\documentclass[10pt,twocolumn,twoside]{IEEEtran}
\usepackage{amsmath,amssymb,amsfonts,amsthm, mathrsfs, bm}% Typical maths resource packages
\usepackage{cite}
\usepackage{array}
\usepackage{enumerate}
\usepackage{graphicx}
\usepackage{url}
\usepackage{color}
\usepackage{algorithm,algorithmic}
\usepackage{multirow}
\usepackage{breqn}
\usepackage[table]{xcolor}
\usepackage[style=base]{caption}
\usepackage[hidelinks]{hyperref}

\graphicspath{{./Figures/}}

\definecolor{gray90}{gray}{0.9}

\newtheorem{Theorem}{Theorem}
\newtheorem{Corollary}[Theorem]{Corollary}
\newtheorem{Proposition}[Theorem]{Proposition}
\newtheorem{Lemma}[Theorem]{Lemma}

\newtheorem{Assumption}{Assumption}

\newtheorem{Theorem_A}{Theorem}[section]
\newtheorem{Proposition_A}{Proposition}[section]
\newtheorem{Lemma_A}{Lemma}[section]

\theoremstyle{remark}
\newtheorem{Rem}{Remark}

\newcommand{\Real}{\mathbb{R}}

\newcommand{\ud}{\mathrm{d}}
\newcommand{\E}{\mathbb{E}}

\renewcommand{\P}{\mathbb{P}}

\newcommand{\cE}{\mathcal{E}}
\newcommand{\oLambda}{{\overline{\Lambda}}}

\newcommand{\ofrac}[1]{{\frac{1}{#1}}}
\newcommand{\ddfrac}[2]{{\frac{\ud #1}{\ud #2}}}

\newcommand{\floor}[1]{{\lfloor {#1} \rfloor}}

\newcommand{\N}[2]{{\mathcal{N}\left(#1,\ #2\right)}}

\newcommand{\norm}[1]{{\left\lVert #1 \right\rVert}}

\newcommand{\ip}[2]{{\left\langle #1, #2 \right\rangle}}
\newcommand{\trace}[1]{{\textrm{\textnormal{Tr}}\left( #1 \right)}}

\newcommand{\EE}[1]{{\mathbb{E}\left[{#1}\right]}}
\newcommand{\Ec}[2]{{\mathbb{E}_{#1}\left[{#2}\right]}}

\begin{document}

\title{Whose Opinion to follow in Multihypothesis Social Learning? A Large Deviations Perspective}
\author{Wee~Peng~Tay,~\IEEEmembership{Member,~IEEE}%
\thanks{Copyright (c) 2014 IEEE. Personal use of this material is permitted. However, permission to use this material for any other purposes must be obtained from the IEEE by sending a request to pubs-permissions@ieee.org.}
\thanks{This research is supported by the MOE Tier 2 grants MOE2013-T2-2-006 and MOE2014-T2-1-028. The author is with the School of Electrical and Electronic Engineering, Nanyang Technological University, Singapore. E-mail: \texttt{wptay@ntu.edu.sg}}
%\thanks{
%Manuscript received July 17, 2011; revised March 31, 2012; accepted July 10, 2012. This work was supported by the MOE AcRF Tier 1 Grant RG25/10. Preliminary versions of parts of this paper were presented at the International Symposium on Wireless and Pervasive Computing, Feb 2011, and at the IEEE International Conference on Acoustics, Speech and Signal Processing, May 2011.  The author is with the Nanyang Technological University, Singapore. E-mail: \texttt{wptay@ntu.edu.sg}
%}%
%\thanks{Copyright (c) 2012 IEEE}
}

%\markboth{IEEE TRANSACTIONS ON INFORMATION THEORY,~Vol.~, No.~, ~2012}%
%{Tay: The Value of Feedback in Decentralized Detection}

%\markboth
%   {To be submitted... }
%   {Tay \MakeLowercase{\textit{et al.}}: }

% Don't write page number 0 to the cover page.
\maketitle \thispagestyle{empty}

% Put abstract and the paper's body in a new page, page 1.
%\newpage
%\setcounter{page}{1}

%---------------------------------------------------------------------------%
%                           abstract and key words                          %
%---------------------------------------------------------------------------%
\begin{abstract}
We consider a multihypothesis social learning problem in which an agent has access to a set of private observations and chooses an opinion from a set of experts to incorporate into its final decision. To model individual biases, we allow the agent and experts to have general loss functions and possibly different decision spaces. We characterize the loss exponents of both the agent and experts, and provide an asymptotically optimal method for the agent to choose the best expert to follow. We show that up to asymptotic equivalence, the worst loss exponent for the agent is achieved when it adopts the 0-1 loss function, which assigns a loss of 0 if the true hypothesis is declared and a loss of 1 otherwise. We introduce the concept of hypothesis-loss neutrality, and show that if the agent adopts a particular policy that is hypothesis-loss neutral, then it ignores all experts whose decision spaces are smaller than its own. On the other hand, if experts have the same decision space as the agent, then choosing an expert with the same loss function as itself is not necessarily optimal for the agent, which is somewhat counter-intuitive. We derive sufficient conditions for when it is optimal for the agent with 0-1 loss function to choose an expert with the same loss function.
\end{abstract}


\section{Introduction}\label{sect:Introduction}

In an increasingly connected world, our opinions on a phenomenon of interest or event are often not only influenced by our direct independent observations, but also by other people's public opinions on related events. In an online social network like Twitter or Facebook, users' opinions and postings are often influenced by the opinions of those they are connected to or are ``following'' in the social network \cite{AraWal:12,MitKilGot:13,Bakshy2011,LuoTayLen:J13}. For example, in viral marketing using social networks, marketing companies often target a few influential nodes in the network to help them push a product \cite{Leskovec2007,Zhang2010}. Similarly, widespread online access has made it easier for us to follow the opinions of experts like celebrities and industry insiders, who may have access to private information that we are unaware of. For example, we may be interested to determine the financial health of a publicly listed company. In addition to our own observations about the company through its annual financial reports and stock prices, we may also choose to incorporate the ``expert'' opinion provided by financial blogs like \cite{ZeroHedge} and \cite{DealBook}. In all these examples, inference about a phenomenon of interest is not only based on direct observations but also the opinions of other entities. This is known as \emph{social learning} \cite{AceDahLobOzd:11,KanTam:13}.

%<*tag:IoT>
A further example is the Internet of Things (IoT) framework \cite{Kortuem2010,Ding2013}. Sensors each make their own private observations but collaborate by exchanging public information. This can be viewed as a ``physical social network'' of devices, cooperating to improve their situational awareness. Sensors originally designed for a specific purpose may collaborate with other sensors to perform inference on a phenomenon they were not specifically designed for. However, in order to ensure energy efficiency, each sensor needs to intelligently choose which other sensors to collaborate with since not all sensors may provide information relevant to it. For example, a sensor trying to estimate the temperature in a particular room of a building may choose to incorporate information from other temperature sensors or sensors tracking the number of occupants in the building, instead of information from a vibration monitoring sensor. 
%</tag:IoT>

%<*tag:setup>
In this paper, we investigate the problem of multihypothesis social learning in which an agent can select an expert opinion from a group of experts, to incorporate into its final decision in order to minimize its expected loss. 
%For example, in inferring the financial health of a publicly listed company, the agent may incorporate the opinion of a financial commentator or professional investment analyst in order to make the best decision for itself. 
Specifically, we consider an agent $0$ who wishes to choose from a set of $M$ hypotheses based on its own observations as well as the opinion of an expert chosen from a set of $K$ possible experts. Each expert has access to a set of private observations, which they use to form their own opinions about the true hypothesis, subject to their own local loss functions or biases, which may differ from that of agent $0$. By taking into account the experts' individual biases, our goal is to find an asymptotically optimal expert choice in order to minimize agent $0$'s expected loss. 
%</tag:setup>

\subsection{Related Work}

The problem of selecting the best expert opinion to follow is related to the problem of \emph{decentralized detection} or decentralized hypothesis testing, which has been extensively studied in \cite{Tsi:93a,CheVar:02,ChaVee:03,LinCheVar:05,WilSwaBlu:00,CheCheVar:12,TayTsiWin:J08b,TayTsiWin:J08a,TayTsiWin:J09a,ZhaPezMor:12,Tay:J12} and the references therein. In the decentralized detection problem, each agent has its own private observations, but cooperate with each other so that the whole network of agents reaches a decision regarding a common underlying phenomenon of interest. Here, which agent passes information to which other agent is determined by a known network topology like the parallel configuration \cite{Tsi:93}, tandem network \cite{PapAth:92,TayTsiWin:J08c,DraOzdTsi:13}, and tree architectures \cite{TayTsiWin:J08a,TayTsiWin:J08b,TayTsiWin:J09a,ZhaPezMor:12}. Therefore, all these works do not consider the problem of selecting which other agent's opinion to follow. Furthermore, in a decentralized detection problem, all agents are assumed to have the same hypotheses and loss functions for declaring the wrong hypothesis, or have the common goal of minimizing the loss at a last agent known as the fusion center.  

We study a related but somewhat different problem from the decentralized detection problem. We consider the scenario where experts may make decisions according to their own biases, instead of minimizing the loss of a particular agent or fusion center. In \cite{AceDahLobOzd:11}, the authors consider binary hypothesis testing in a social network, in which agents sequentially observe the opinions of a stochastically generated neighborhood of agents in the network. Each agent has the same 0-1 loss function,\footnote{The 0-1 loss function assigns a loss of 0 if the agent declares the same hypothesis as the true underlying hypothesis, and a loss of 1 otherwise.} and conditions are derived for the Bayesian error probability of the $n$th agent to approach to zero as $n$ becomes large. The network model we consider in this paper is equivalent to a two-layer hierarchical tree network, which is much simpler compared to that studied in \cite{AceDahLobOzd:11}. This is because our goal is to analyze how an expert's opinion impacts that of a particular agent. In addition, we consider a multihypothesis testing problem in which each agent has different loss functions or even different number of hypotheses. 

%<*tag:NP>
Since finding optimal decision rules for general decentralized hypothesis testing problems is NP-complete \cite{TsiAth:85}, it is difficult to find analytical characterizations for the best expert choice in networks of moderate size, and where agents have private observations that are not conditionally independent given the underlying hypothesis. Therefore, most of the literature in decentralized detection has focused on the case where observations are conditionally independent \cite{Tsi:93a,CheVar:02,ChaVee:03,LinCheVar:05,WilSwaBlu:00,CheCheVar:12,TayTsiWin:J08b,TayTsiWin:J08a,TayTsiWin:J09a,ZhaPezMor:12}. The same challenge applies to the social learning problem in this paper, and we therefore consider only the case where agents' private observations are conditionally independent. Although the problem becomes numerically tractable under this assumption, it remains hard to characterize the agents' optimal policies and decision rules analytically (see \cite{TayTsiWin:J09a} for a discussion).  To overcome this difficulty, we consider the regime where each agent has access to an asymptotically large number of private observations (e.g., over a sufficiently long period of time) but each agent's observation sources and loss functions may differ. This allows us to adopt a large deviations perspective to the expert choice problem, and derive analytical characterizations of the optimal choice and agent strategies. 
%</tag:NP>

%<*tag:NitAtiVee>
The reference \cite{NitAtiVee:13} considers a multihypothesis testing problem in a parallel configuration with a single agent, which is allowed to take actions that affect the observations it makes. The distribution of the observation at each time step depends on the action of the agent in the previous time step, and has bounded Kullback-Liebler divergences under any pair of hypotheses. The agent's aim is to minimize the error exponent corresponding to the 0-1 loss function. Our work can be viewed as a generalization of a result in \cite{NitAtiVee:13} (which also considers the sequential detection problem that we do not study here) to the case where one of the agent observations is the opinion of an expert, which itself has asymptotically many private observations leading to unbounded Kullback-Liebler divergences for its opinion. Furthermore, we consider general loss functions and the corresponding loss exponent, which is a generalization of the error exponent in \cite{NitAtiVee:13}. 
%</tag:NitAtiVee>
The characterization of the error exponents corresponding to 0-1 loss functions in multihypothesis testing has also been studied in \cite{Tun:05,GriHar:10}, which derive the achievable error exponents region. 

\subsection{Our Contributions}\label{subsection:contributions}

Our goal is to optimize the loss exponent of agent $0$, the absolute value of which is the rate of decay of the expected loss incurred by agent $0$ as the number of private observations grows large. We suppose that agent $0$ has a decision space consisting of $M$ hypotheses, and has a choice of $K$ expert opinions to follow.
\begin{enumerate}[(i)]
	\item We consider a general $M$-ary multihypothesis social learning framework in which an agent $0$ and every expert have loss functions that depend on the number of private observations the agent or expert has access to. Each expert makes a decision that takes values from a decision space, the size of which may not be the same as $M$. We characterize the loss exponent of each expert, and provide an asymptotically optimal policy for an expert to achieve its optimal loss exponent (Theorem \ref{theorem:expert}).
	\item We derive the optimal loss exponent for agent $0$ after it has incorporated the opinion of a particular expert, and provide an asymptotically optimal method for agent $0$ to choose the best expert to follow (Theorem \ref{theorem:expert_choice}).  We also show that the worst loss exponent, up to asymptotic equivalence,\footnote{See Section \ref{subsection:asymptotic} for the definition of asymptotic equivalence.} for agent $0$ is achieved when agent $0$ adopts the 0-1 loss function (cf.\ Remark \ref{rem:worstloss} of Section \ref{subsect:choose}).
		\item We introduce the concept of hypothesis-loss neutrality in Section \ref{subsection:special}, and show that if agent $0$ adopts a hypothesis-loss neutral policy, then the opinion of any expert who has a decision space with number of states strictly less than $M$, is ignored by agent $0$ (Proposition \ref{prop:expert_necessary}). In this case, additional information is useless, which is somewhat unexpected.  
	\item We show that if all experts have the same decision space as agent $0$, then it is not necessarily optimal for agent $0$ to choose the expert with the same loss function as itself. This is surprising as conventional wisdom seems to suggest otherwise. If agent $0$ adopts the 0-1 loss function, we derive sufficient conditions for when it is optimal to choose an expert who also utilizes the 0-1 loss function (Proposition \ref{prop:bestexpertloss}).  

\end{enumerate}

%<*tag:game>
In this work, we do not address the case where agent $0$'s opinion may be utilized by one of the experts, which result in much more complex opinion dynamics than that considered in this paper. This leads to the interesting question of whether there exists an equilibrium in the choice of loss functions in a network of agents who can incorporate opinions from each other. This is however out of the scope of the current work, and will be addressed in our future research.
%</tag:game>

The rest of this paper is organized as follows. In Section \ref{sect:Formulation}, we describe our system model, problem formulation and assumptions. We characterize the agents' loss exponents in Section \ref{sect:Optimal}, and provide asymptotically optimal policies to achieve the optimal loss exponents. In Section \ref{sect:OptimalChoice}, we discuss the optimal expert choice for agent $0$, and derive insights into this choice by making simplifying assumptions. We conclude in Section \ref{sect:Conclusion}. Appendix \ref{appendix:Mathematical} contains a brief review of some basic definitions of large deviations theory, and we defer all proofs to Appendix \ref{appendix:proofs}. Appendix \ref{appendix:Ak_properties} contains a characterization of the asymptotic decision regions, which are introduced in Section \ref{subsect:lossexp_experts}.

\section{Problem Formulation}\label{sect:Formulation}

In this section, we define our system model, assumptions and some notations. We consider an underlying measurable space $(\Omega,\mathcal{F})$ on which all random variables in this paper are defined. We adopt the following notations throughout this paper. Let $\Real$ be the space of real numbers, and let $\ip{t}{z}$ be the inner product of the vectors $t$ and $z$ in $\Real^{M-1}$. For $z \in \Real^{M-1}$, let $z^0 = (0,z) \in \Real^M$ be the vector augmented with a zero as the first element. The range of integers $a,a+1,\ldots,b$ is denoted as $[a,b]$. The notations $x[1:n]$ and $(x_i)_{i=1}^n$ are used to represent the sequences $x=(x[1],x[2],\ldots,x[n])$ and $(x_1, x_2, \ldots, x_n)$, respectively. We also make use of $x[i]$ to denote the $i$th element in the vector or sequence $x$. 

\subsection{Learning From an Expert}\label{subsect:Learning_Network}

\begin{figure}[!tb]
\begin{center}
\includegraphics[scale=1]{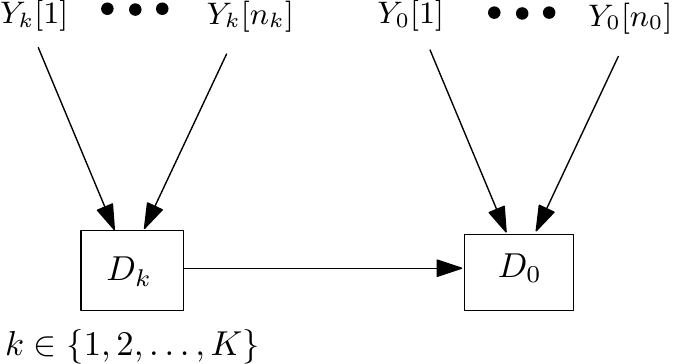}
\caption{Learning network in which agent $0$ incorporates the decision of an agent $k \in\{1,2,\ldots,K\}$.
}\label{fig:two_layer}
\end{center}
\end{figure}

Suppose that an agent $0$ wishes to determine the underlying hypothesis $H$ associated with a phenomenon of interest, which from agent $0$'s frame of reference can be modeled by a set of probability measures $\{\P_m : m = 0,1,\ldots,M-1\}$ on the space $(\Omega,\mathcal{F})$. Let $\E_m$ be the mathematical expectation under $\P_m$, and let $H = m \in [0,M-1]$ if all agents' private observations have distributions derived from the probability measure $\P_m$. We suppose that $H=m$ has prior probability $\pi_m \in (0,1)$. The agent $0$ can choose to incorporate the opinion of an expert, chosen from a set of expert agents $\{1,2,\ldots,K\}$ (see Figure \ref{fig:two_layer}). If there is no need to distinguish between agent $0$ and the experts, we use the generic term ``agent'' to refer to either of them.

%<*tag:conditionalprob>
Each agent $k \in \{0,\ldots,K\}$ has access to a set of private observations $Y_k[1:n_k] = (Y_k[1],\ldots,\allowbreak Y_k[n_k])$. Conditioned on $H=m$, we assume that each $Y_k[l]$, for $l=1,\ldots,n_k$, is drawn independently from a conditional distribution belonging to the set $\{\P_m^\gamma: \gamma \in \Gamma_k\}$, where $\Gamma_k$ is a finite index set representing the information sources of agent $k$. The distribution set $\{\P_m^\gamma: \gamma \in \Gamma_k\}$ varies from agent to agent, and models the differences in quality of information that each agent may have access to. 
%<*tag:knowP>
 We assume that for each $\gamma\in\Gamma_k$, the probability measures $\P_m^\gamma$ for $m=0,\ldots,M-1$, are absolutely continuous with respect to (w.r.t.) each other, and are known to all the agents. 
%</tag:knowP>
 We further assume that conditioned on $H$, all the random variables $(Y_k[1:n_k])_{k=0}^K$ are independent.\footnote{Although two experts may have the same information source, we assume their private observations from this source are independent due to their own noisy interpretations of the same piece of information.}
%</tag:conditionalprob>

%<*tag:motivationex>
To motivate our setup, consider the example alluded to in Section \ref{sect:Introduction}, where an agent $0$ is interested to determine the financial health of a large publicly listed conglomerate. In addition to its own observations $Y_0[1:n_0]$ about the company through annual financial reports and other publicly available indicators like stock prices, the agent $0$ may also choose to incorporate the expert opinion of an investment analyst. Because of limited financial resources (since most analyst reports are not free), agent $0$ can only choose to subscribe to one analyst, and has to make an optimal choice of which analyst to use. In most of this paper, for simplicity, we restrict ourselves to the case where the opinion of a single expert is considered in the decision making of agent $0$, and show how to generalize our results to a finite set of experts later in Remark \ref{remark:multiexperts} of Section \ref{sect:OptimalChoice}. Note also that the number of information sources for each agent is also typically finite as in this example, which justifies our assumption that $|\Gamma_k| < \infty$ for all $k\geq 0$. 
%</tag:motivationex>

%<*tag:xkdef>
Let $x_k = (x_k[\gamma])_{\gamma\in\Gamma_k}$ be a vector of non-negative weights summing to one, where $\floor{x_k[\gamma]n_k}$ is the number of private observations of agent $k$ that have conditional distribution $\P_m^\gamma$ when the hypothesis $H=m$. We let the remaining $n_k - \sum_{\gamma\in\Gamma_k}\floor{x_k[\gamma]n_k}$ observations be drawn from an arbitrary distribution from the set $\{\P_m^\gamma: \gamma \in \Gamma_k\}$. Since we are concerned about asymptotics in this paper, the choice of the arbitrary distribution is immaterial to our analysis. We say that $x_k$ is a \emph{policy} for agent $k$. An agent chooses its policy in order to minimize its expected loss, as described below. 
%</tag:xkdef>
Let $\mathbb{S}(\Gamma_k)$ be the simplex consisting of all agent policies. 

%<*tag:policychoice>
Based on its observations, each expert $k \geq1$ makes a decision $D_k = \gamma_k(Y_k[1:n_k]) \in [0,d_k-1]$ by minimizing a local loss criterion. We assume that every expert $k \geq 1$ chooses its policy $x_k$ and decision rule $\gamma_k$ so that 
\begin{align}\label{Ck}
\EE{C_k(H, D_k, n_k)}
\end{align}
is minimized, where $C_k(m,d,n_k)$ is a non-negative finite loss incurred if the decision of expert $k$ is $d$ when the true hypothesis is $H=m$. 
%</tag:policychoice>
 We can think of $C_k(\cdot,\cdot,\cdot)$ as encoding the ``bias'' of expert $k$. We call this the \emph{loss function} of expert $k$,
%\footnote{To be technically correct, $(C_k(\cdot,\cdot,n_k))_{n_k\geq1}$ should be called a sequence of loss functions, but for simplicity, we will not adopt this terminology.} 
and
\begin{align}\label{exp_k}
\limsup_{n_k \to \infty} \ofrac{n_k} \log \EE{C_k(H, D_k, n_k)}
\end{align}
the \emph{loss exponent} of expert $k$. An example of a loss function is the 0-1 loss function: $C_k(m,m,n_k) = 0$, and $C_k(m,d,n_k)=1$ for all $n_k\geq 1$ if $d \ne m$, which results in \eqref{Ck} being the error probability considered in \cite{Tsi:88}. In this paper, we consider the regime where experts have asymptotically large number $n_k$ of private observations, which corresponds to the case where experts are very experienced in their respective fields (as determined by $\Gamma_k$).

%<*tag:lossnk>
By allowing the loss function to depend on the number of private observations, we can model various practical applications. For example, the expert $k$ may itself be part of a decentralized detection network like a tree configuration \cite{TayTsiWin:J08c,TayTsiWin:J09a}, in which case its goal is to minimize the Bayesian error probability at a fusion center. Then, its loss function decays exponentially fast in the number of private observations $n_k$. In the publicly listed conglomerate example, suppose that one of the hypotheses corresponds to the conglomerate being financially bankrupt. The loss associated with a missed detection of this hypothesis can be modeled to be exponentially larger than the loss associated with a missed detection of another more benign hypothesis.
%</tag:lossnk>

In our model, expert $k$'s decision space consists of $d_k$ states, where $d_k$ may not equal to $M$, the number of hypotheses that agent $0$ is interested in. This allows us to model scenarios where different agents may have different models for the underlying state of the world. Using the conglomerate example described above, agent $0$ may be interested in the financial health of the whole conglomerate, while a particular analyst may only be interested in the real estate arm of the conglomerate. Nevertheless, the performance of the real estate arm has a bearing on the overall health of the conglomerate. In particular, expert $k$ may not distinguish between two hypotheses, say $H=M-2$ and $H=M-1$. This can be modeled by assuming that the expert declares the decision $d=m$ when $H=m < M-2$ and $d=M-2$ if $H=M-2$ or $M-1$. In this case, we let $C_k(M-2,d,n_k)=C_k(M-1,d,n_k)=1$ for all $d<M-2$ and $C_k(M-2,M-2,n_k)=C_k(M-1,M-2,n_k)=0$. This model is also general enough to model different decision spaces that may not map directly to any of the hypotheses of agent $0$. 

On the other hand, we note that without loss of generality, there is no need to consider the case where an expert uses a model in which there are more than $M$ hypotheses, since agent $0$ can always choose a $M$ that is sufficiently large in its model. In addition, since our analysis is based on the frame of reference of agent $0$, it has no knowledge of any additional hypotheses, and in practical applications, it simply assumes that expert $k$ minimizes the expected loss given in \eqref{Ck}. We assume that based on the publicized expertise of each expert, agent $0$ knows its loss function decay rates, defined in Assumption \ref{assumpt:loss} below.  

\subsection{Loss of Agent 0}

Let $D_0(k) = \gamma_0(Y_0[1:n_0],D_k) \in[0,M-1]$ be the decision made by agent $0$ after incorporating the opinion $D_k$ of agent $k$. The expected loss of agent $0$ is  
\begin{align*}
\EE{C_0(H, D_0(k), n_0)},
\end{align*}
where $C_0(m, d, n_0)$ is the non-negative loss incurred by agent $0$ if it decides in favor of hypothesis $d$ when the true hypothesis is $H=m$, and the number of private observations is $n_0$. 
We make the following assumptions regarding the loss function of each agent.
\begin{Assumption}\label{assumpt:loss}\
\begin{enumerate}[(i)]
	\item\label{it:c_0} For all $n_0\geq1$ and $m \in [0,M-1]$, we have $C_0(m,m,n_0)=0$, and for every $d \in [0,M-1]$ such that $d \ne m$, we have $0< C_0(m,d,n_0) \leq 1$ with
	\begin{align*}
	\hspace{-8pt}-\hspace{-6pt}\lim_{n_0\to\infty} \hspace{-2pt}\ofrac{n_0}\log C_0(m,d,n_0) = c_0(m,d) \equiv c_0(m) \in [0,\infty).
	\end{align*}
	
	\item\label{it:c_k} For each agent $k \geq 1$, we have for every $m \in [0,M-1]$ and $d\in [0,d_k-1]$,
	\begin{align*}
	-\lim_{n_k\to\infty} \ofrac{n_k}\log C_k(m,d,n_k) = 	c_k(m,d) \geq 0.
	\end{align*}
	We assume that agent $0$ knows the loss function decay rates $c_k(\cdot,\cdot)$, for all $k=1,\ldots,K$, based on the publicized expertise of each expert. 			
\end{enumerate}
\end{Assumption}

Since the number of agents and decision states are finite, we can normalize $C_k(m,d,n_k)$ by $\displaystyle\max_{m,d,k} C_k(m,d,n_k)$ or $\displaystyle\sum_{d,k} C_k(m,d,n_k)$ so that there is no loss in generality in assuming that $C_k(m,d,n_k) \leq 1$ for all $m$, $d$, and $k$. In Assumption~\ref{assumpt:loss}\eqref{it:c_0}, we make the simplifying assumption that the loss function decay rate for agent $0$ w.r.t.\ a particular hypothesis is the same for all wrong decisions. This is because otherwise, easy examples can be constructed in which agent $0$ declares the wrong hypothesis with high probability for large $n_0$. 
%<*tag:exampleAssumption1>
To see this, suppose that all hypotheses have the same prior probability, $H=0$ is the true hypothesis, and agent $0$'s observations are all drawn from the same conditional distribution. The loss of agent $0$ if it makes the decision $d$ is, with high probability, approximately proportional to
\begin{align*}
\sum_{m\ne d} e^{-n_0 (c_0(m,d) + K_{0m})},
\end{align*}
where $K_{0m}$ is the Kullback-Liebler divergence of the conditional distribution under $H=0$ versus that under $H=m$ \cite{CovTho:B05}. If $c_0(0,1) > c_0(1,0)+K_{01}$, while $c_0(m,1)>c_0(m,0)$ for all $m\ne 0,1$, then agent $0$ decides in favor of $H=1$ instead of $H=0$ when $n_0$ is large. This is clearly an undesirable model as $c_0(m,1) > c_0(m,0)$ for all $m \ne 0,1$ implies that the agent imposes an exponentially smaller loss when declaring $H=1$ versus $H=0$. In this case, an arguably more appropriate modeling approach is to merge the hypothesis $H=1$ into another hypothesis. 
%</tag:exampleAssumption1>

In Assumption~\ref{assumpt:loss}\eqref{it:c_k}, we assume that agent $0$ knows the loss function decay rates of the experts. This assumption may not hold in some practical applications, in which case we need to impose uncertainties on agent $0$'s knowledge about the experts, similar in spirit to \cite{HoTayQuek:C14}. This unfortunately makes the problem much more challenging, and is out of the scope of our current work. The results in this work serve as the basic foundation on which the more challenging problem in which agent $0$ has limited knowledge of the experts, can be addressed in future research.

\subsection{Asymptotic Equivalence and Optimal Loss Exponent}\label{subsection:asymptotic}

Consider two loss functions with loss decay rates $c_k(m,d)$ and $c_k'(m,d)$ respectively. If $c_k(m,d)$ and $c_k'(m,d)$ differ by the same constant for all $m$ and $d$, then their corresponding loss exponents \eqref{exp_k} are different but the optimal policies to minimize \eqref{exp_k} are the same. We therefore say that two loss functions are \emph{asymptotically equivalent} if their respective loss decay rates $c_k(m,d)$ and $c_k'(m,d)$ differ by the same constant for all $m$ and $d$. It can be shown that each loss function belongs to an equivalence class, with the equivalence relation being asymptotic equivalence. We call a loss function with $\min_{m,d} c_k(m,d) = 0$ a \emph{canonical loss function} of its equivalence class. For example, the 0-1 loss function is a canonical loss function of its equivalence class. In another example, consider the set of all loss functions $C_k(m,d,n_k)$ with $\sum_d C_k(m,d,n_k) = C$ for each $m$ and $n_k$, which imposes a total loss of $C$ for each hypothesis. Then, it can be shown that each of these loss functions is a canonical loss function.

We say that two loss exponents are equivalent if they have the same policy and their loss functions are asymptotically equivalent. For fair comparison of loss exponents, we will always assume canonical loss functions. We are interested to characterize the optimal loss exponent\footnote{Note that because of Assumption \ref{assumpt:loss}, the loss exponent is \emph{negative}, with a more negative loss exponent corresponding to a faster loss decay rate.}
\begin{align}\label{loss_exponent}
\min_{1\leq k\leq K}
\limsup_{n_0\to\infty} \ofrac{n_0}\log \EE{C_0(H, D_0(k), n_0)},
\end{align}
when the number of private observations of agent $0$ becomes large. Note that although agent $0$ is allowed an asymptotically large number of private observations, the quality of information available to agent $0$ is constrained by the set $\Gamma_0$. The experts' information sources $\{\Gamma_k:k=1,\ldots,K\}$ are different from $\Gamma_0$, and may thus improve the loss exponent of agent $0$. We assume that for all experts $k\geq1$, $q_k = \lim_{n_0\to\infty} n_k /n_0$ exists as a limit. If the number $n_k$ of private observations of expert $k$ is such that $q_k = 0$, agent $0$ will ignore the opinion of expert $k$ as its opinion becomes asymptotically negligible compared to agent $0$'s private observations. Therefore, without loss of generality, we assume that $q_k > 0$ for all $k \geq 1$.

\subsection{Technical Definitions}\label{subsect:Notations}

%<*tag:technicaldef>
In this subsection, we define notations that will be commonly used throughout the paper. We rely heavily on the mechanisms of large deviations theory to characterize the agents' loss exponents \eqref{exp_k} and \eqref{loss_exponent}. We refer the reader to Appendix \ref{appendix:Mathematical} for a brief overview of some basic concepts, and to \cite{DemZei:98} for a detailed treatment of large deviations theory. As the theory of large deviations in hypothesis testing problems utilizes log moment generating functions of log likelihood ratios, and their Fenchel-Legendre transforms in order to characterize the loss exponents, we define these necessary quantities in the following. 
%</tag:technicaldef>

For any given random variable $X$ with marginal distribution $\P_i^X$ under hypothesis $H=i$, we abuse notation by letting $\ell_{ij}(X)$ be the Radon-Nikodym derivative (or likelihood ratio) of $\P_i^X$ w.r.t.\ $\P_j^X$. Note that $\ell_{ij}(X)$ is a random variable that depends on the distributions and realization of $X$. 
%For a particular realization of the random variable $X=x$, we use $\ell_{ij}(X=x)$ to denote the realization of the corresponding Radon-Nikodym derivative. 
By convention, we let $\ell_{ii}(X) = 1$ for all values of $i$ and all realizations of $X$. In addition, for simplicity,  we let $\ell_{ij}^\gamma = \ud\P_i^\gamma/\ud\P_j^\gamma$ be the Radon-Nikodym derivative  of $\P_i^\gamma$ w.r.t.\ $\P_j^\gamma$, for each $\gamma\in\bigcup_k \Gamma_k$. 

Let $Z^\gamma = (\log \ell_{m0}^\gamma)_{m=1}^{M-1}$ be a vector of log likelihood ratios, and the log moment generating function of $Z^\gamma$ under $H=m$ be
\begin{align*}
\xi_{m}(\gamma,t) = \log \Ec{m}{\exp(\ip{t}{Z^{\gamma}})},
\end{align*}
for all $t \in \Real^{M-1}$. For a policy $x = (x[\gamma])_{\gamma\in\Gamma_k}$ of an agent $k$, the weighted log moment generating function is then given by
\begin{align}\label{logmomentgen}
\varphi_m(t,x) = \sum_{\gamma\in\Gamma_k} x[\gamma] \xi_m(\gamma,t),
\end{align}
and its Fenchel-Legendre transform \cite{DemZei:98} is
\begin{align}\label{Fenchel}
\Phi_m^*(z,x) = \sup_{t\in\Real^{M-1}} \left\{ \ip{t}{z} - \varphi_m(t,x)\right\},
\end{align}
where $z \in \Real^{M-1}$. The function $\Phi_m^*(z,x)$ will be shown in Theorem \ref{theorem:expert} to characterize the rates of decay of the probabilities $\P_m(D_k=d)$, similar to the rate functions in large deviations theory \cite{DemZei:98}.

When an expert $k$'s decision space has size $d_k=M$, the rates of decay of the probabilities $\P_m(D_k=d)$ become easier to characterize since we can now interpret a decision $d$ of the expert $k$ to be in favor of hypothesis $d$. Then, under the true hypothesis $H=m$, the log moment generating function of interest is the one involving the likelihood ratios of the observation distributions under hypothesis $d$ versus hypothesis $m$, given by
\begin{align}\label{def:Lambdaij}
\Lambda_{ij}(s,x) = \sum_{\gamma\in\Gamma_k} x[\gamma] \log \Ec{i}{(\ell_{ji}^\gamma)^s}.
\end{align}
The Fenchel-Legendre transform of $\Lambda_{ij}$ is then given by
\begin{align*}
\Lambda_{ij}^*(z,x) = \sup_{s\in\Real} \left\{ sz - \Lambda_{ij}(s,x)\right\},
\end{align*}
where $z\in\Real$.

We assume that the distributions $\{\P_m^\gamma$, $\gamma\in\bigcup_k\Gamma_k\}$ are well-behaved for all $m\in [0,M-1]$ in the following assumption, which holds for example in the case where all conditional distributions are from the exponential families. This assumption is required in all our proofs in order to show that an agent's loss exponent cannot be better than a certain achievable lower bound, which we characterize in order to derive the agent's optimal policy.

\begin{Assumption}\label{assumption:xi}
For all $m\in [0,M-1]$ and all $\gamma\in\bigcup_k\Gamma_k$, $\xi_m(\gamma,t) < \infty$ for all $t\in\Real^{M-1}$.
\end{Assumption}

\subsection{Results Overview}\label{subsection:results}
%<*tag:resultsoverview>
If agent $0$ incorporates the opinion of expert $k$, we expect the loss exponent of expert $k$ to determine how useful its opinion is to agent $0$. Therefore, we first find the loss exponent of each expert $k$ in Theorem \ref{theorem:expert}, which also provides an asymptotically optimal policy for expert $k$. Then, we proceed to determine the loss exponent of agent $0$ if it incorporates the opinion of a particular expert $k$ in Theorem \ref{theorem:agent0}. To find the optimal expert, we simply optimize the loss exponent found in Theorem \ref{theorem:agent0} over the whole set of experts, which is essentially the content of Theorem \ref{theorem:expert_choice}. Based on the conclusions of Theorem \ref{theorem:expert_choice}, we provide detailed procedures in Remark \ref{rem:hardness} of Section \ref{subsect:choose} that allow agent $0$ to compute its asymptotically optimal expert choice and policy. Finally, we provide insights into the optimal expert choice under simplifying assumptions in Propositions \ref{prop:expert_necessary} and \ref{prop:bestexpertloss}.  
%</tag:resultsoverview>

\section{Optimal Loss Exponents}\label{sect:Optimal}

In this section, we first characterize the loss exponents of the experts $k=1,2,\ldots,K$, which leads to an asymptotically optimal policy for each expert. We then characterize the loss exponent of agent $0$, assuming that it is following the opinion of some expert $k$. 

\subsection{Loss Exponents of Experts}\label{subsect:lossexp_experts}

Consider an expert $k$, where $k =1,\ldots,K$. By conditioning on the observations $Y_k[1:n_k]$, it can be shown (Proposition 2.3 of \cite{Tsi:93}) that the optimal decision rule for expert $k$ is given by
\begin{align}\label{optimal_D}
D_k = \arg \min_{0\leq d \leq d_k-1} \sum_{m=0}^{M-1} \pi_m C_k(m,d,n_k)\ell_{m0}(Y_k[1:n_k]),
\end{align}
where $\ell_{m0}(Y_k[1:n_k]) = \prod_{l=1}^{n_k} \ell_{m0}(Y_k[l])$. In general, the right hand side of \eqref{optimal_D} may have multiple minimizers. 
%<*tag:uniquemin>
In order to avoid having to consider the use of randomization to determine the final decision for agent $k$ (due to a mathematical technicality in the proof of Theorem \ref{theorem:expert} below), and for the ease of interpreting results, we will assume throughout this paper that the right hand side of \eqref{optimal_D} has a unique solution with probability one. 
%</tag:uniquemin>

\begin{Assumption}\label{assumpt:unambiguous}
For every agent $k=1,\ldots,K$, and for any policy $x_k$, the minimization on the right hand side of \eqref{optimal_D} has a unique solution with probability one.
\end{Assumption}

%<*tag:uniquemincond>
Assumption \ref{assumpt:unambiguous} is satisfied if conditioned on any hypothesis $H \in \{0,\ldots,M-1\}$, 
%the joint conditional probability distribution of $(\ell_{m0}^\gamma)_{m\geq 1}$, for any $\gamma\in\Gamma_k$, is absolutely continuous w.r.t.\ Lebesgue measure, i.e., 
$(\ell_{m0}^\gamma)_{m\geq 1}$ are continuous random variables with a joint probability density function. To see this, it can be shown via an easy inductive argument that $(\ell_{m0}(Y_k[1:n_k]))_{m\geq 1}$ are in turn continuous random variables under any hypothesis $H$. Assumption \ref{assumpt:unambiguous} then follows by the same argument in Lemma 5.1 of \cite{Tsi:93a}, which we refer the reader to. This shows that Assumption \ref{assumpt:unambiguous} holds in most practical applications as private observations are usually modeled as noisy, with the noise typically having a probability density function like the Gaussian probability density \cite{You09, Viv96}.
%</tag:uniquemincond>

If agent $0$ follows expert $k$, its loss exponent depends on the loss exponent \eqref{exp_k} of the expert $k$, which is in turn related to the probability exponents
\begin{align}\label{exp_k_md}
\limsup_{n_k \to \infty} \ofrac{n_k} \log \P_m(D_k = d),
\end{align}
where $m \in [0,M-1]$ and $d \in [0,d_k-1]$. In the following, our aim is to characterize the probability exponents \eqref{exp_k_md}. We use \eqref{optimal_D} to characterize \eqref{exp_k_md} by first letting 
\begin{align*}
\tilde{g}_k(z, d, n_k) &= \ofrac{n_k} \log \sum_{m=0}^{M-1} \pi_m e^{\log C_k(m,d,n_k) + n_kz[m]}, 
\end{align*}
and
\begin{align*}
g_k(z,d,n_k) &= \tilde{g}_k(z,d,n_k) - \min_{d' \ne d} \tilde{g}_k(z,d',n_k),
\end{align*}
where $z = (z[0], z[1], \ldots, z[M-1]) \in \Real^{M}$. Suppose that agent $k$ adopts the policy $x_{k,n_k}$ when it has access to $n_k$ private observations. Then from \eqref{optimal_D} and Assumption \ref{assumpt:unambiguous}, we obtain
\begin{align}\label{probD}
\P_m(D_k=d) = \P_m(g_k(\bar{Z}_{n_k}^0(x_{k,n_k}), d, n_k) < 0),
\end{align}
where 
\begin{align}\label{barZ}
\bar{Z}_{n_k}(x_{k,n_k}) = \left( \ofrac{n_k} \log \ell_{m0}(Y_k[1:n_k]) \right)_{m=1}^{M-1}
\end{align}
is a vector in $\Real^M$ of log likelihood ratios, and $\bar{Z}_{n_k}^0(x_{k,n_k}) = (0, \bar{Z}_{n_k}(x_{k,n_k}))$ is the vector of log likelihood ratios augmented with $0$ as the first entry (this corresponds to $1/n_k \cdot \log \ell_{00}(Y_k[1:n_k]) = 0$). In the following, we show that $g_k(z,d,n_k)$ converges uniformly in $z$ to
\begin{align}\label{fk}
f_k(z,d) &= \tilde{f}_k(z,d) - \min_{d' \ne d} \tilde{f}_k(z,d'),
\end{align}
where
\begin{align}\label{tfk}
\tilde{f}_k(z, d) &= \max_{0\leq m \leq M-1}\{z[m] - c_k(m,d) \}.
\end{align}

\begin{Lemma}\label{lemma:f_uniform}
Suppose that Assumption \ref{assumpt:loss} holds. For all $k=1,\ldots,K$, and all $d \in [0,d_k-1]$, $\tilde{g}_k(z, d, n) \to \tilde{f}_k(z, d)$ and $g_k(z,d,n) \to f_k(z,d)$ uniformly in $z \in \Real^{M}$ as $n\to\infty$.
\end{Lemma}
\begin{IEEEproof}
See Appendix \ref{appendix:proofs}.
\end{IEEEproof}

We are now ready to present our first main result. Since we are working in the regime of large $n_k$, Lemma \ref{lemma:f_uniform} tells us that we can replace $g_k$ with $f_k$ in \eqref{probD}. For $k \geq 1$, let 
\begin{align}\label{Ak}
A_k(d) = \left\{z \in \Real^{M-1}: f_k(z^0, d) < 0,\textrm{ where } z^0 = (0, z)\right\},
\end{align}
The set $A_k(d)$ can be interpreted as the asymptotic decision region for expert $k$ to declare decision $d$ based on its sufficient statistics $\bar{Z}_{n_k}(x_{k,n_k})$; see Section \ref{subsect:ADR} for a discussion. Assumption \ref{assumpt:unambiguous} is required here to ensure that the sets $A_k(d), d\in[0,d_k-1]$ are disjoint. 

\begin{Theorem}\label{theorem:expert}
Suppose that Assumptions \ref{assumpt:loss}, \ref{assumption:xi}, and \ref{assumpt:unambiguous} hold. Consider an agent $k \in \{1,\ldots, K\}$ who adopts a sequence of optimal policies $(x_{k,n_k})_{n_k \geq 1}$ that minimizes \eqref{Ck} for each $n_k$. Then, we have the following:
\begin{enumerate}[(i)]
	\item\label{it:expert_opt} The loss exponent of agent $k$ is given by 
	\begin{align*}
	\lim_{n_k\to\infty} \ofrac{n_k} \log \EE{C_k(H, D_k, n_k)} = - \max_{x \in \mathbb{S}(\Gamma_k)} I_k(x),
	\end{align*}
	where
	\begin{align}\label{Ik}
	&\hspace{-8pt}I_k(x) =\hspace{-8pt} \min_{\substack{0\leq m \leq M-1 \\ 0 \leq d \leq d_k-1}} \left\{\inf_{z \in A_k(d)} \Phi_m^*(z, x) + c_k(m,d)\right\}.
	\end{align}
	
	\item\label{it:policy_opt} There is no loss in optimality asymptotically, if we restrict the sequence of policies $(x_{k,n_k})_{n_k \geq 1}$ such that $\lim_{n_k \to\infty} x_{k,n_k} = x_k^*$, for some $x_k^* \in \arg\max_{x \in \mathbb{S}(\Gamma_k)} I_k(x)$.
	
	\item\label{it:prob_exp} Let $x_k^* \in \arg\max_{x \in \mathbb{S}(\Gamma_k)} I_k(x)$. For every $m\in [0,M-1]$, and $d \in [0,d_k-1]$, we have
	\begin{align*}
	\lim_{n_k \to \infty} \ofrac{n_k}\log\P_m(D_k = d) = - \inf_{z \in A_k(d)} \Phi_m^*(z, x_k^*).
	\end{align*}
\end{enumerate}
\end{Theorem}
\begin{IEEEproof}
See Appendix \ref{appendix:proofs}.
\end{IEEEproof}

Theorem \ref{theorem:expert} allows each expert to find an asymptotically optimal policy by maximizing \eqref{Ik} (see Remark \ref{rem:hardness} in Section \ref{subsect:choose}). In addition, we will see later that Theorem \ref{theorem:expert}\eqref{it:prob_exp} allows agent $0$ to characterize its own loss exponent and thus optimally choose the expert to follow.

\begin{Rem}\label{rem:Ik}
%<*tag:rem:Ik>
The quantity $I_k(x)$ in \eqref{Ik} can be interpreted as the rate of loss decay for agent $k$ adopting policy $x$, i.e., for $n_k$ sufficiently large, we have
\begin{align*}
\EE{C_k(H, D_k, n_k)} \approx g(n_k) e^{-I_k(x)}, 
\end{align*}
 where $g(n_k)$ is a function that decays faster than exponentially in $n_k$. To achieve a smaller loss, agent $k$ then chooses a policy $x$ that maximizes $I_k(x)$. Observe that $I_k(x)$ is the minimization over all $m\in[0,M-1]$ and $d\in[0,d_k-1]$ of the terms $\inf_{z \in A_k(d)}\Phi_m^*(z, x) + c_k(m,d)$, where $\inf_{z \in A_k(d)}\Phi_m^*(z, x)$ is the absolute probability exponent in Theorem \ref{theorem:expert}\eqref{it:prob_exp}, while $c_k(m,d)$ is the loss function decay rate of declaring decision $d$ when the true hypothesis is $m$. Therefore, each term in the minimization of the right hand side of \eqref{Ik} is the absolute decay rate of $C_k(m,d,n_k)\P_m(D_k=d)$, and the overall loss exponent is dominated by the term that decays the slowest, which is what we expect from large deviations theory (cf.\ Lemma 1.2.15 of \cite{DemZei:98}). We next provide an intuitive interpretation for the regions $A_k(d)$ in the following subsection.
%</tag:rem:Ik>
\end{Rem}

\subsection{Asymptotic Decision Regions}\label{subsect:ADR}

Suppose that agent $k\geq1$ adopts the sequence of policies $x_{k,n_k}\to x_k$ as $n_k\to\infty$. From \eqref{probD}, for large $n_k$, we have for each $m\in [0,M-1]$, and $d \in [0,d_k-1]$,
\begin{align*}
\ofrac{n_k}\log \P_m(D_k=d) 
&\approx  \ofrac{n_k}\log \P_m(f_k(\bar{Z}_{n_k}(x_{k,n_k}), d) < 0) \\
&\approx - \inf_{z \in A_k(d)} \Phi_m^*(z, x_k),
\end{align*}
where $A_k(d)$ is as defined in \eqref{Ak}, and $\Phi_m^*(z, x_k)$ can be interpreted as a rate function. In the same spirit as the G\"artner-Ellis Theorem \cite{DemZei:98}, we can interpret the sets $A_k(d)$ as the asymptotic decision region for deciding $D_k = d$ based on $\bar{Z}_{n_k}(x_{k,n_k})$ as $n_k\to\infty$, where the rate of decay of $\P_m(D_k=d)$ is dominated by the rate at a particular realization of $\bar{Z}_{n_k}(x_{k,n_k})$ in the region $A_k(d)$. We refer the reader to Appendix \ref{appendix:Ak_properties} for a characterization of $A_k(d)$ in terms of the union of intersections of multiple half-spaces. 

In the following, we list some properties of the rate functions $\Phi_m^*(\cdot,\cdot)$ that will be useful in helping us to interpret our results. 

\begin{Lemma}\label{lemma:Phi_properties}
For every $m\in[0,M-1]$, $\Phi_m^*(z,x)$ is non-negative, convex in $z$ and concave in $x$. Furthermore, for any policy $x$, $\displaystyle\min_{z\in\Real^{M-1}} \Phi_m^*(z,x) = 0$, and the minimum is achieved at 
\begin{align}\label{eqn:tildez}
\tilde{z}_m(x) = \sum_{\gamma} x[\gamma] \Ec{m}{Z^\gamma},
\end{align}
where $Z^\gamma = (\log \ell_{i0}^\gamma)_{i=1}^{M-1}$.
\end{Lemma}
\begin{IEEEproof}
See Appendix \ref{appendix:proofs}.
\end{IEEEproof}

In particular, if $d_k=M$, and the loss function of agent $k$ adopting the policy $x_k$ is such that $\tilde{z}_m(x_k) \in A_k(m)$ in \eqref{eqn:tildez} for every $m$, then we can interpret the decision $m$ of agent $k$ as in favor of hypothesis $H=m$ since $\P_m(D_k=m)$ is bounded away from 0 and has decay rate $\inf_{z \in A_k(m)} \Phi_m^*(z, x_k) = 0$.

\subsection{Loss Exponent of Agent 0}

In the following, we characterize the loss exponent of agent $0$ if it chooses expert $k$. Recall that $q_k = \lim_{n_0\to\infty} n_k /n_0$.

\begin{Theorem}\label{theorem:agent0}
Suppose that Assumptions \ref{assumpt:loss}, \ref{assumption:xi}, and \ref{assumpt:unambiguous} hold. Suppose that agent $0$ adopts the opinion of expert $k\geq 1$, which has the asymptotically optimal policy $x_k^*$. Then, the loss exponent of agent $0$ is
\begin{align*}
&\lim_{n_0\to\infty} \ofrac{n_0}\log \EE{C_0(H,D_0(k),n_0)} = - \max_{x_0\in\mathbb{S}(\Gamma_0)} \cE_0(k,x_0),
\end{align*}
where \footnote{To avoid cluttered notations, we let $\min_{i\ne j}$ and $\max_{i\ne j}$ be the minimization or maximization over all unordered pairs $(i,j) \in [0,M-1]^2$ such that $i\ne j$, respectively.}
\begin{align}\label{C0_lowerbound}
&\cE_0(k,x_0) \nonumber\\
&= \min_{\substack{i\ne j \\ 0\leq d \leq d_k-1}} \max_{s\in [0,1]} \bigg\{  (1-s)\left(q_k\inf_{z \in A_k(d)} \Phi_i^*(z, x_k^*) + c_0(i)\right) \nonumber\\
&\quad\quad + s\left(q_k\inf_{z \in A_k(d)} \Phi_j^*(z, x_k^*) + c_0(j)\right) - \Lambda_{ij}(s,x_0) \bigg\}.
\end{align}
\end{Theorem}
\begin{IEEEproof}
See Appendix \ref{appendix:proofs}.
\end{IEEEproof}

Following Remark \ref{rem:Ik}, we can again interpret each term in the minimization on the right hand side of \eqref{C0_lowerbound} as the absolute error exponent incurred when differentiating between hypotheses $i$ and $j$. From Theorem \ref{theorem:expert} and Theorem \ref{theorem:agent0}, we see that the choice of an expert $k$ affects agent $0$'s loss exponent through the rate of decay of the probabilities $\P_i(D_k=d)$ and $\P_j(D_k=d)$, which is to be expected. In particular, if the loss function of the expert $k$ is such that we can interpret the decision $d$ of the expert as in favor of hypothesis $d$, then $\P_i(D_k=i) > 0$ yields $\inf_{z \in A_k(d)} \Phi_i^*(z, x_k^*)=0$, and it does not contribute to the decay of the expected loss of agent 0. 
%If we set $c_0(i) = 0$ for all $i\in[0,M-1]$, and $q_k=0$, we recover the same error exponent for the decentralized hypothesis testing problem in \cite{Tsi:88}.

\section{Optimal Choice of Expert}\label{sect:OptimalChoice}

We now address the question of how to choose an optimal expert for agent $0$. We first revisit Theorem \ref{theorem:agent0} to derive the optimal loss exponent for agent $0$, and then we make additional simplifying assumptions in order to provide more insights into our results. Finally, we discuss a numerical example to illustrate the use of our optimal loss exponent characterizations in finding the optimal policies and expert choice.

\subsection{Choosing an Expert}\label{subsect:choose}

If agent $0$ does not incorporate the opinion of any expert, we use the notation $\cE_0(0,x_0)$ to denote its absolute loss exponent when using policy $x_0$. From Theorem \ref{theorem:agent0}, by setting $q_k=0$, we have
\begin{align}\label{E0}
\cE_0(0,x_0) &= \min_{i\ne j} \max_{s\in [0,1]} \left\{  
(1-s)c_0(i) + sc_0(j) - \Lambda_{ij}(s,x_0) \right\}.
\end{align}
In the case of minimizing the error probability at agent $0$ with 0-1 loss function, and without the help of any experts, we can further set $c_0(i)=c_0(j)=0$ to obtain the absolute error exponent 
\begin{align}\label{EB}
\cE_{0,B}(0,x_0) &= -\max_{i\ne j} \min_{s\in [0,1]} \Lambda_{ij}(s,x_0),
\end{align}
which recovers the result in \cite{Tsi:88}. Similarly, if agent $0$ chooses expert $k$, and adopts the 0-1 loss function, we let its absolute loss exponent be
\begin{dmath}[label={EkB}]
\cE_{0,B}(k,x_0) = \min_{\substack{i\ne j \\ 0\leq d \leq d_k-1}} \max_{s\in [0,1]} \left\{  (1-s)q_k\inf_{z \in A_k(d)} \Phi_i^*(z, x_k^*) + sq_k\inf_{z \in A_k(d)} \Phi_j^*(z, x_k^*) - \Lambda_{ij}(s,x_0) \right\}.
\end{dmath}

The following proposition follows immediately from Theorem \ref{theorem:agent0}, and provides a method for agent $0$ to optimally choose which expert to follow.

\begin{Theorem}\label{theorem:expert_choice}
Suppose that Assumptions \ref{assumpt:loss}, \ref{assumption:xi}, and \ref{assumpt:unambiguous} hold. The optimal loss exponent of agent $0$ is 
\begin{dmath*}
\min_{1\leq k\leq K}\lim_{n_0\to\infty} \ofrac{n_0}\log \EE{C_0(H,D_0(k),n_0)} = -\max_{1\leq k \leq K} \max_{x_0\in\mathbb{S}(\Gamma_0)} \cE_0(k,x_0).
\end{dmath*}
Furthermore, for any $k\geq 1$ and policy $x_0$, we have
\begin{align}
&\cE_0(k,x_0) \geq \cE_{0,B}(k,x_0) \label{Eineq1}\\ 
&\cE_0(k,x_0) \geq \cE_0(0,x_0) \geq \cE_{0,B}(0,x_0). \label{Eineq2}
\end{align}
\end{Theorem}
\begin{IEEEproof}
See Appendix \ref{appendix:proofs}.
\end{IEEEproof}
\begin{Rem}\label{rem:hardness}
Theorem \ref{theorem:expert_choice} shows that agent $0$ should choose an expert $k$ that maximizes $\max_{x_0} \cE_0(k,x_0)$, which depends on $\inf_{z \in A_k(d)} \Phi_m^*(z, x_k^*)$, for $m\in[0,M-1]$ and $d\in[0,d_k-1]$. We now discuss procedures that can achieve this, depending on the amount of information that agent $0$ has about the experts:
\begin{enumerate}[(i)]
	\item\label{it:probexp} If the experts' probability exponents $\{\inf_{z \in A_k(d)} \Phi_m^*(z, x_k^*): m\in[0,M-1],d\in[0,d_k-1]\}$ for all $k \geq 1$ are publicized, then agent $0$ simply needs to find its optimal policy $x_0$ that maximizes $\cE_0(k,x_0)$ in Theorem \ref{theorem:expert_choice} for each $k$, and choose the expert that produces the largest $\max_{x_0} \cE_0(k,x_0)$. To find the optimal policy $x_0$ corresponding to an expert $k$, we can use an iterative procedure as follows:
		\begin{enumerate}[Step 1.]
		\item An initial guess for the optimal policy $x_0^*(0)$ is made. Set $l=1$.
		\item\label{step:s} For each pair $(i,j)\in[0,M-1]$, $i\ne j$, and $d\in[0,d_k-1]$, find $s_{ijd}(l)$ that is the maximizer of the optimization problem over $s\in[0,1]$ on the right hand side of \eqref{C0_lowerbound}. Note that this is equivalent to a convex minimization problem, which can be solved via standard convex optimization methods \cite{BoyVan:B04}. 		
		\item\label{step:x0} Find $x_0^*(l)=(x[\gamma])_{\gamma\in\Gamma_0}$ by solving the linear program:
		\begin{align*}
		& \max_{r, (x[\gamma])_{\gamma\in\Gamma_0}} \quad r \\
		& \textrm{subject to } \\
		& r \leq (1-s_{ijd}(l))\left(q_k\inf_{z \in A_k(d)} \Phi_i^*(z, x_k^*) + c_0(i)\right) \\
		&\quad\quad + s_{ijd}(l) \left(q_k\inf_{z \in A_k(d)} \Phi_j^*(z, x_k^*) + c_0(j)\right) \\		
		& \qquad - \sum_{\gamma\in\Gamma_0} x[\gamma] \log \Ec{i}{(\ell_{ji}^\gamma)^{s_{ijd}}} ,\\
		& \qquad \forall i,j \in[0,M-1], i\ne j, \forall d\in[0,d_k-1], \\
		& x[\gamma] \geq 0, \ \forall \gamma\in\Gamma_0,\\
		& \sum_{\gamma\in\Gamma_0} x[\gamma] = 1.
		\end{align*}		
		\item Set $l = l+1$ and repeat Steps 2-4 till $x_0^*(l)$ does not change significantly.
	\end{enumerate}

	\item\label{it:lossandpolicy} If agent $0$ knows only the experts' loss functions and optimal policies $x_k^*$ for all $k\geq 1$, then it can compute $\inf_{z \in A_k(d)} \Phi_m^*(z, x_k^*)$ for all $m\in[0,M-1]$ and $d\in[0,d_k-1]$ by searching for the minimum of $\Phi_m^*(z,x_k^*)$ on the boundaries of $A_k(d)$, if $\tilde{z}_m(x_k^*)$ in \eqref{eqn:tildez} is not in $A_k(d)$. This is because $\Phi_m^*(z,x_k^*)$ is convex in $z$ (cf. Lemma \ref{lemma:Phi_properties}), and the search can be performed using standard convex optimization methods \cite{BoyVan:B04}. The boundaries of $A_k(d)$ can be found through its characterization in Appendix \ref{appendix:Ak_properties} (see Section \ref{subsection:numerical} for an example). If $\tilde{z}_m(x_k^*)\in A_k(d)$, then from Lemma \ref{lemma:Phi_properties}, we obtain $\inf_{z \in A_k(d)} \Phi_m^*(z, x_k^*)=0$. Once the experts' probability exponents have been computed, the same procedure as in item \eqref{it:probexp} above can now be used to choose the optimal expert.
	
		\item\label{it:lossonly} If agent $0$ knows only the experts' loss functions, it can utilize Theorem \ref{theorem:expert}\eqref{it:expert_opt} to determine the policy that each expert adopts. This however may not be an easy numerical procedure if $M$ and $d_k$ are large, even in the case where the expert has 0-1 loss function \cite{Tsi:88}. We propose the following alternating optimization approach for finding the optimal policy of expert $k$:
	\begin{enumerate}[Step 1.]
		\item An initial guess for the optimal policy $x_k^*(0)$ is made. Set $l=1$.

		\item\label{step:t} For each $m\in[0,M-1]$ and $d\in[0,d_k-1]$, find $z_{m,d}(l) = \arg\inf_{z\in A_k(d)} \Phi_m^*(z,x_k^*(l-1))$. As noted in item \eqref{it:lossandpolicy} above, this can be obtained via standard convex optimization methods. Let 
		\begin{dmath*}
		t_{m,d}(l)) \hiderel{=} \arg\max_{t\in\Real^{M-1}} \left\{ \ip{t}{z_{m,d}(l-1)} - \varphi_m(t,x_k^*(l))\right\}.
		\end{dmath*}
		
		\item\label{step:x} Find $x_k^*(l)=(x[\gamma])_{\gamma\in\Gamma_k}$ by solving the linear program
		\begin{align*}
		& \max_{r, (x[\gamma])_{\gamma\in\Gamma_k}} \quad r \\
		& \textrm{subject to } \\
		& r \leq \ip{t_{m,d}(l)}{z_{m,d}(l)} \\
		& \qquad - \sum_{\gamma\in\Gamma_k} x[\gamma] \xi_m(\gamma,t_{m,d}(l)) + c_k(m,d),\\
		& \qquad \forall m\in[0,M-1], \forall d\in[0,d_k-1], \\
		& x[\gamma] \geq 0, \ \forall \gamma\in\Gamma_k,\\
		& \sum_{\gamma\in\Gamma_k} x[\gamma] = 1.
		\end{align*}
		
		\item Set $l = l+1$ and repeat Steps 2-4 till $x_k^*(l)$ does not change significantly.
	\end{enumerate}
Unfortunately, there is no guarantee that the above procedure converges to the correct solution. However, in our numerical experiments, we were able to arrive at the correct optimal policy by using multiple initial guesses. A numerical example is presented in Section \ref{subsection:numerical}. 
		
\end{enumerate}

\end{Rem}

\begin{Rem}\label{rem:worstloss}
The first inequality in \eqref{Eineq2} shows that there is no loss in optimality for agent $0$ to incorporate the opinion of any expert, verifying the adage that there is no harm in having more information. However, incorporating additional information does not necessarily improve its loss exponent; see Proposition \ref{prop:expert_necessary}. 

The inequality in \eqref{Eineq1} shows that of all the loss functions satisfying Assumption \ref{assumpt:loss}\eqref{it:c_0}, the 0-1 loss is the worst canonical loss function for agent $0$. In particular, consider the set of loss functions with $C_0(m,d,n_0)=C_0(m,n_0)$ for all $m\in[0,M-1]$ and $d \ne m$, where the limit $c_0(m) = -\lim_{n_0\to\infty} (1/n_0)\allowbreak\log C_0(m,n_0)$ exists. Furthermore, each loss function has a total cost constraint,
\begin{align*}
\sum_{m=0}^{M-1} C_0(m,n_0) = M-1.
\end{align*}
It can be shown that all such loss functions satisfy Assumption \ref{assumpt:loss}\eqref{it:c_0}, and are canonical loss functions. The 0-1 loss function belongs to this set, and divides the total cost equally among all types of missed detections, which results in the worst loss exponent for agent $0$. This can be explained intuitively by observing that if there exists a $m$ such that $c_0(m) > 0$, then missed detection of $H=m$ incurs an exponentially decaying loss, so that the hypothesis $m$ can effectively be ignored. Agent $0$ then effectively has a smaller set of hypotheses, leading to a lower expected loss.
\end{Rem}

%\begin{Rem}\label{rem:ignore}
%Suppose that agent $0$ adopts the policy $x_0$, and incorporates the opinion of agent $k\geq 1$ with corresponding asymptotically optimal policy $x_k^*$. Suppose further that $(i^*, j^*)$ is such that 
	%\begin{align*}
	%\max_{s\in [0,1]} \left\{(1-s)c_0(i^*) + sc_0(j^*) - \Lambda_{ij}(s,x_0) \right\} = \cE_0(0,x_0),
	%\end{align*}
%i.e., the loss of agent $0$ is dominated by its error in distinguishing between hypotheses $i^*$ and $j^*$. Then, equality in the first inequality of \eqref{Eineq2} holds if there exists a $d \in [0,d_k-1]$ such that $\inf_{z \in A_k(d)} \Phi_{i^*}^*(z, x_k^*) = \inf_{z \in A_k(d)} \Phi_{j^*}^*(z, x_k^*) = 0$. The asymptotic acceptance region corresponding to $d$ thus includes the minima of the rate functions corresponding to hypotheses $i^*$ and $j^*$. This implies that expert $k$ cannot distinguish between these two hypotheses with probability of error going to zero exponentially fast in the number of private observations, and is thus useless to agent $0$, as verified by Theorem \ref{theorem:agent0}.
%\end{Rem}

\begin{Rem}\label{remark:multiexperts}
It is easy to generalize Theorem \ref{theorem:agent0} or Theorem \ref{theorem:expert_choice} to the case where more than one agent can be chosen. In particular, if all $K$ experts' opinions are adopted by agent $0$, Theorem \ref{theorem:agent0} holds with $\cE_0(k,x_0)$ replaced by
\begin{align*}
&\min_{\substack{i\ne j \\ (p_k)_{k=1}^K}}\hspace{-8pt} \max_{s\in [0,1]} \bigg\{  (1-s)\left(\sum_{k=1}^K q_k \inf_{z \in A_k(p_k)} \Phi_i^*(z, x_k^*) + c_0(i)\right) \nonumber\\
&\quad\quad + s\left(\sum_{k=1}^K q_k\inf_{z \in A_k(p_k)} \Phi_j^*(z, x_k^*) + c_0(j)\right) - \Lambda_{ij}(s,x_0) \bigg\}
\end{align*}
where the sequences $(p_k)_{k=1}^K \in \prod_{k=1}^K [0,d_k-1]$.
\end{Rem}

%Since the optimal expert choice depends on how well agent $0$ differentiates its hypotheses without the help of any expert, not much more can be said regarding the expert choice due to the generality of the assumptions used in Theorem \ref{theorem:expert_choice}. 
In the following subsection, we consider the special cases where the size of each expert's decision space is less than or equal to $M$ under additional simplifying assumptions.

\subsection{Special Cases}\label{subsection:special}

In this section, we assume that agent $0$'s policy has been fixed in advance. We first consider the case where an expert $k$ may have a decision space smaller than the number of hypotheses $M$ that agent $0$ has. To state our results, suppose that agent $0$ adopts policy $x_0$ and has loss decay rates given by the function $c_0$. We define the \emph{loss augmented} log moment generating function for $i,j\in[0,M-1]$ as 
\begin{align}\label{eqn:oLambda}
\oLambda_{ij}(s,x_0,c_0) = \Lambda_{ij}(s,x_0) - (1-s)c_0(i) - sc_0(j).
\end{align}
For the case where agent $0$ adopts the 0-1 loss function, we have $c_0(i)=0$ for all $i$. We observe that in this case, the Chernoff information $-\min_{s\in[0,1]}\Lambda_{ij}(s,x_0)$ tells us how well hypothesis $i$ can be differentiated from hypothesis $j$, where a larger Chernoff information corresponds to a faster error probability decay \cite{DemZei:98}. This leads us to the following general definition: For a given policy $x_0$, if there exists a constant $\alpha$ so that $\min_{s\in[0,1]}\oLambda_{ij}(s,x_0,c_0) = \alpha$ for all $i\ne j$, we say that the policy $x_0$ is \emph{hypothesis-loss neutral} for agent $0$. 

Intuitively, a policy is hypothesis-loss neutral if it can differentiate any pair of hypotheses equally well when adjusted for the agent's ``biases''. It may be argued that in practical scenarios, an agent would strive to be hypothesis-loss neutral, since otherwise there are hypotheses that are relatively less important than others, and can be dropped. In the following, we show that if agent $0$ is hypothesis-loss neutral, an expert's opinion is useless if it does not have a decision space as large as $M$.

\begin{Proposition}\label{prop:expert_necessary}
Suppose that Assumptions \ref{assumpt:loss}, \ref{assumption:xi}, and \ref{assumpt:unambiguous} hold. Suppose that agent $0$ adopts a policy $x_0$ that is hypothesis-loss neutral. Then, for any agent $k\geq 1$, if $d_k < M$, we have $\cE_0(k,x_0) = \cE_{0,B}(0,x_0)$, i.e., agent $0$ ignores the opinion of agent $k$.
\end{Proposition}
\begin{IEEEproof}
See Appendix \ref{appendix:proofs}.
\end{IEEEproof}

Proposition \ref{prop:expert_necessary} can be explained intuitively as follows in the case where agent 0 has 0-1 loss function. If an expert $k$ has $d_k < M$, then there exists two hypotheses $i$ and $j$ of agent $0$ that it does not discriminate between. The probability of making an error between these two hypotheses by the expert is therefore one, and the expert's opinion does not help agent $0$ to differentiate between hypotheses $i$ and $j$. The error probability of agent $0$ declaring $H=i$ when the true hypothesis is $H=j$ or vice versa, thus dominates when agent $0$ is hypothesis-loss neutral, and is the same as when agent $0$ ignores the expert's opinion.

We next consider the case where experts have the same decision space as agent $0$, and has zero loss if they decide on the true underlying hypothesis. We also make the following simplifying assumption.
\begin{Assumption}\label{assumpt:Bayesian2}
Every agent $k \geq1$ has $d_k = M$, with loss decay rates
\begin{align*}
c_k(m,d) = \left\{
\begin{array}{ll}
c_k(m) & \textrm{ if $d \ne m$,}\\
\infty & \textrm{ if $d = m$,},
\end{array}
\right.
\end{align*}
where $c_k(m) \geq 0$ for all $m\in[0,M-1]$. 
\end{Assumption}

\begin{Proposition}\label{prop:bestexpertloss}
Suppose that Assumptions \ref{assumpt:loss}-\ref{assumpt:Bayesian2} hold. Suppose that agent $0$ adopts policy $x_0$, and follows the opinion of expert $k \geq 1$, who adopts the policy $x_k$. Then, the loss exponent of agent $0$ is 
\begin{align}\label{sp_bestlossexp}
&\lim_{n_0\to\infty} \ofrac{n_0}\log \EE{C_0(H,D_0(k),n_0)} = - \tilde\cE_0(k,x_0),
\end{align}
where \footnote{The notation $\min_{i,j: i\ne j}$ means minimization over all \emph{ordered} pairs $(i,j)$ with $i,j\in[0,M-1]$ and $i\ne j$.}
\begin{dmath*}
\tilde{\cE}_{0}(k,x_0) =  \min_{i,j: i\ne j}\max_{s\in [0,1]} \left\{ sq_k \Lambda^*_{ji}(c_k(i)-c_k(j),x_k) - \oLambda_{ij}(s,x_0,c_0) \right\}.
\end{dmath*}
In addition, if for some $i\ne j$ such that $c_0(i)=c_0(j)$, and 
\begin{align}\label{01condition}
\Lambda_{ij}(s,x_0) = \Lambda_{ji}(s,x_0),\quad \forall s\in [0,1],
\end{align}
then there is no loss in optimality for agent 0 to restrict to experts $k$ with canonical loss functions satisfying $c_k(i)=c_k(j)$.
\end{Proposition}
\begin{IEEEproof}
See Appendix \ref{appendix:proofs}.
\end{IEEEproof}

Surprisingly, Proposition \ref{prop:bestexpertloss} shows that it is not necessarily optimal for agent $0$ to choose an expert who utilizes the same canonical loss function as itself since agent $0$'s loss exponent depends only on the relative differences in agent $k$'s loss function at each hypothesis. We show a numerical example of this phenomenon in Section \ref{subsection:numerical} below. In particular, if agent $0$ adopts the 0-1 loss function, and \eqref{01condition} does not hold for some $i\ne j$, then an example can be constructed in which an expert with a loss function different from the 0-1 loss is optimal for agent $0$. 

On the other hand, if agent $0$ is ``unbiased'' (in terms of loss) towards a pair of hypotheses $(i,j)$ and \eqref{01condition} holds, Proposition \ref{prop:bestexpertloss} tells us that it is optimal for agent $0$ to choose an expert, if any, who is also ``unbiased'' towards hypotheses $(i,j)$. This indicates that if agent $0$ has the same discriminatory power for a particular pair of hypotheses (which is the intuitive meaning of \eqref{01condition}), then it values information from an ``unbiased'' expert more than one who has the same ``bias'' as itself. Assuming that independent news agencies are in general ``unbiased,'' our result suggests that such news agencies are unlikely to be replaced by social news reporting or social blogs. The following corollary follows immediately from Proposition \ref{prop:bestexpertloss}.
\begin{Corollary}\label{Cor:01loss}
Suppose that Assumptions \ref{assumpt:loss}-\ref{assumpt:Bayesian2} hold. Suppose that agent $0$ adopts policy $x_0$, and has 0-1 loss function. If for every distinct pair of hypotheses $(i,j)$, \eqref{01condition} holds, then it is optimal for agent $0$ to choose an expert, if any, who also has the 0-1 loss function.
\end{Corollary}

\subsection{Numerical Example}\label{subsection:numerical}

In this section, we present a numerical example to illustrate our results. Suppose that $M=3$, there are 3 experts to choose from, and all agents have access to private observations from two information sources $\Gamma = \{\gamma_1, \gamma_2\}$, whose conditional distributions are shown in Table \ref{table:information_sources}. We use the notation $\N{\mu}{\sigma^2}$ to denote the normal distribution with mean $\mu$ and variance $\sigma^2$.\footnote{Normal distributions have been used to model social opinions and influences in the economics literature, including \cite{You09, Viv96}.} For example, the three hypotheses can correspond to the financial health of a company being ``bad'', ``neutral'', and ``good'' respectively. The information source $\gamma_1$ represents a view that tends to be optimistic when the true health of the company is ``neutral'', while $\gamma_2$ represents a view that tends to be pessimistic.

\begin{table}[!htb]
	\centering
		\caption{Conditional distributions of the information sources.}
		\begin{tabular}{ccc}
			\hline
														Hypothesis $H$ 	& $\gamma_1$             & $\gamma_2$ 			\\
			\hline 									
			 											$0$				      & $\N{-1}{\sigma^2}$      & $\N{-1}{\sigma^2}$ 	\\
			\rowcolor{gray90}			$1$				      & $\N{\delta}{\sigma^2}$  & $\N{-\delta}{\sigma^2}$ 	\\	
														$2$				      & $\N{1}{\sigma^2}$       & $\N{1}{\sigma^2}$ 	\\												
			\hline
		\end{tabular}
	\label{table:information_sources}
\end{table}

\begin{figure}[!htb]
\centering
\includegraphics[width=0.4\textwidth]{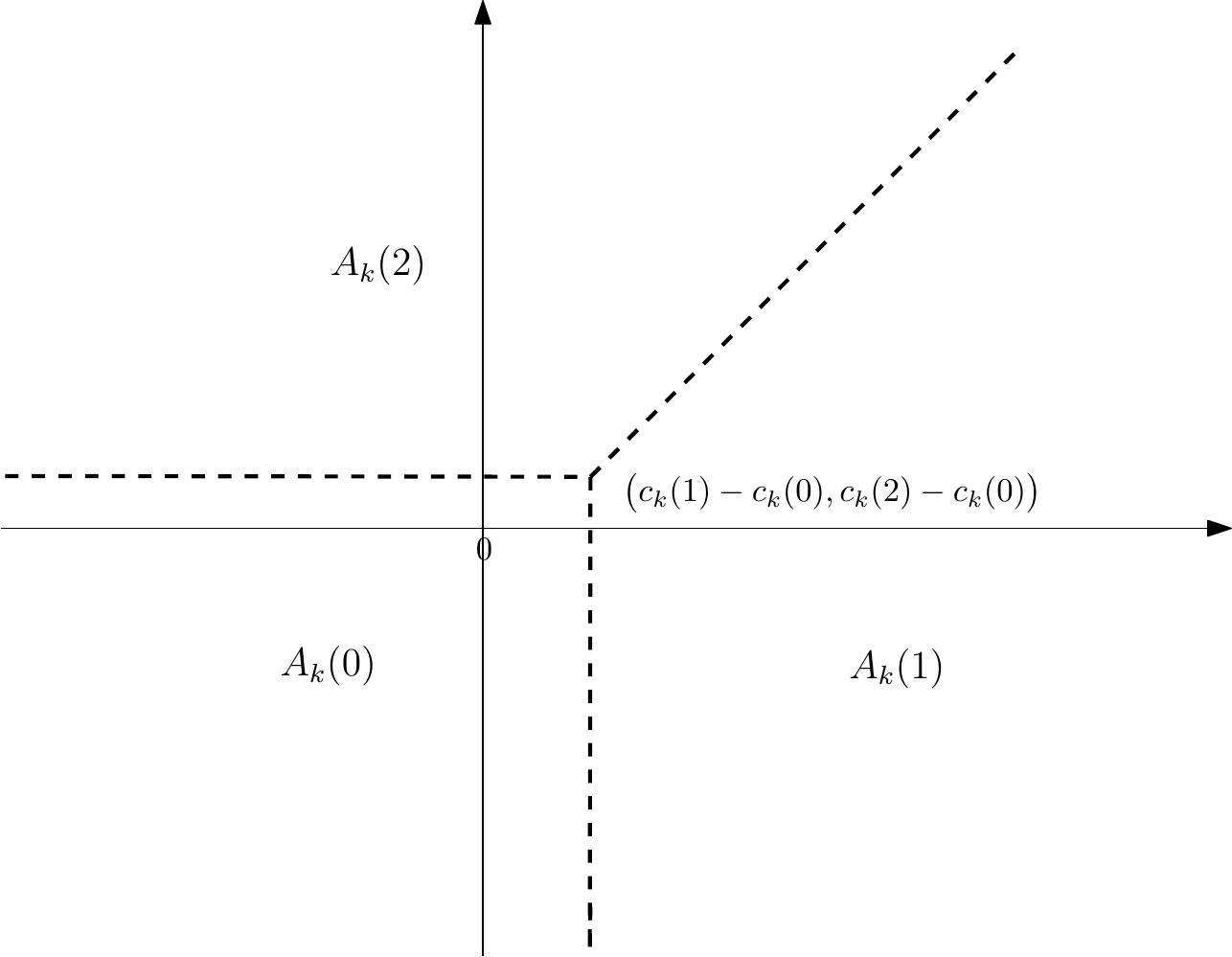}
\caption{Asymptotic decision regions for agent $k$. The dotted lines denote the boundaries between the regions.}\label{Fig:Ak}
\end{figure}

In this example, we take $\sigma=2$, $q_k=1$ for all $k\geq 1$, and assume Assumption \ref{assumpt:Bayesian2} for easier interpretation of our results. The asymptotic decision regions of each agent $k \geq 1$ can be shown, using the characterization in Appendix \ref{appendix:Ak_properties}, to be that in Figure \ref{Fig:Ak}. For each agent $k$, let $c_k = (c_k(0),c_k(1),c_k(2))$ be its vector of loss decay rates. Let $c_1=(0,0,0)$, $c_2=(0,0,0.2)$, and $c_3=(0,0.05,0)$. We use the alternating optimization procedure in Remark \ref{rem:hardness} item \eqref{it:lossonly} to find the optimal policies for each agent. An exhaustive search is done to determine the actual optimal policy for each expert, which is given in Table \ref{table:optimal policies}. Note that the experts' optimal policies are invariant of $\delta$.

\begin{table}[!htb]
	\centering
		\caption{Optimal policies for agents. The leftmost columns show the optimal policies for each expert. The three rightmost columns show the optimal policy and loss decay rate of an agent 0 with $c_0=c_3$, corresponding to each expert choice.}
		\begin{tabular}{ccc|ccc}
			\hline
														Expert $k$	  & $x_k^*[\gamma_1]$   & $x_k^*[\gamma_2]$ 	& $x_0^*[\gamma_1]$   & $x_0^*[\gamma_2]$  & $\cE_0(k,x_0)$		\\
			\hline 									
			 											$1$				& 0.5             & 0.5 	& 0.5   & 0.5 		& 0.1099\\
			\rowcolor{gray90}			$2$				& 1          			& 0 	& 0.2117         	& 0.7883 	& 0.1158\\
														$3$				& 0.5       			& 0.5 	& 0.5       			& 0.5 		& 0.1066\\												
			\hline
		\end{tabular}
	\label{table:optimal policies}
\end{table}

\begin{figure}[!htb]
\centering
\includegraphics[width=0.45\textwidth]{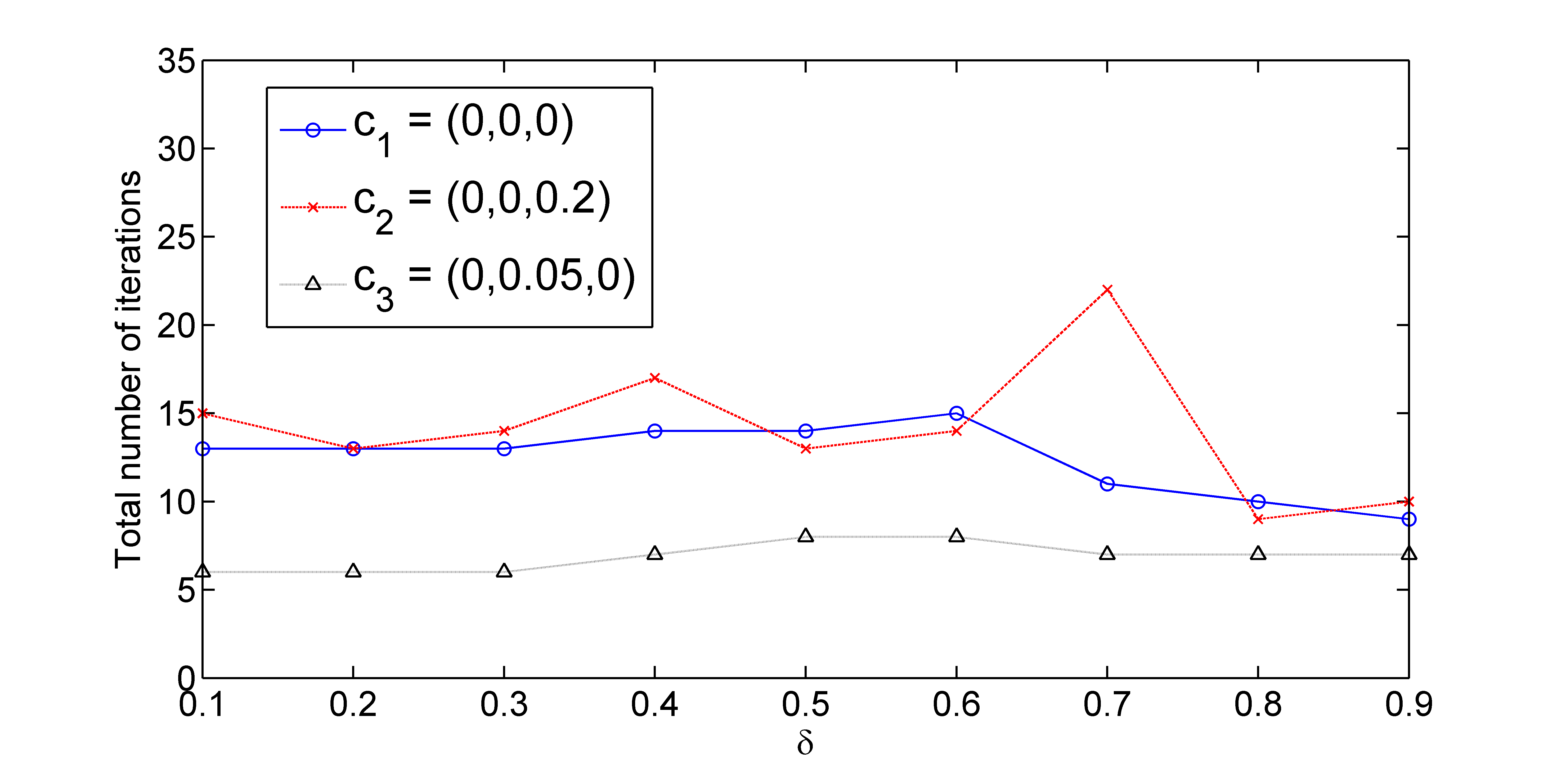}
\caption{Total number of iterations required to find optimal policies for experts 1, 2, and 3.}\label{Fig:delta}
\end{figure}

For experts 1 and 3, their optimal policies are both given by $x_1[\gamma_1]=x_1[\gamma_2]=0.5$, as is expected from the symmetry of the problem. For expert 2, its optimal policy is given by $x_2[\gamma_1]=1$ and $x_2[\gamma_2]=0$. This is because with a positive loss decay rate $c_2(2)=0.2$, expert 2 puts exponentially less loss on confusing hypothesis 2 for the others, therefore it chooses $\gamma_1$ as its sole information source. For the alternating optimization procedure, we started our initial guesses at $(x[\gamma_1] , x[\gamma_2]) = (0,1)$ and $(0.3,0.7)$.\footnote{Although a more natural initial guess is $(0.5,0.5)$, this was not selected in order to test the convergence for experts 1 and 3.} We repeated the procedure for different values of $\delta$. In all cases, we were able to find the correct optimal policies within 22 iterations (totaled over the two runs corresponding to the two initial guesses), as shown in Figure \ref{Fig:delta}. In some cases however, the procedure does not converge to the correct optimal policy if we restrict ourselves to the single initial guess $(x[\gamma_1] , x[\gamma_2]) = (0,1)$. This shows that the proposed procedure can get stuck at a suboptimal policy.

We next suppose that $c_0 = c_3$, and use the procedure in Remark \ref{rem:hardness} item \eqref{it:probexp} to obtain the optimal policy for agent $0$ w.r.t.\ each expert $k=1,2,3$, when $\delta=0.9$. We found the optimal policies for agent $0$ within 4 iterations in each case, and that choosing expert 2 produces the best loss exponent for agent $0$, as shown in Table \ref{table:optimal policies}. In this case, it is not optimal for agent $0$ to choose expert 3, which has the same loss function as itself. Observe that even if agent $0$'s policy is fixed at $x_0=(0.5,0.5)$, it will still prefer expert 1 over expert 3. On the other hand, if $c_0=c_1$, we have $\cE_0(1,x_0)=0.0884 > \cE_0(3,x_0) = 0.0750 > \cE_0(2,x_0) = 0.0566$, i.e., agent $0$'s optimal expert choice is expert 1. It can be shown that \eqref{01condition} holds for all pairs of hypotheses in this example, thus verifying Corollary \ref{Cor:01loss}. 

%\begin{table}[!ht]
	%\centering
		%\caption{Optimal policies for agent 0.}
		%\begin{tabular}{cccc}
			%\hline
														%Expert Choice  & $x_0^*[\gamma_1]$   & $x_0^*[\gamma_2]$  & $\cE_k(k,x_0)$ 			\\
			%\hline 									
			 											%$1$				& 0.5             & 0.5 		& 0.1099\\
			%\rowcolor{gray90}			$2$				& 0.2117         	& 0.7883 	& 0.1158\\	
														%$3$				& 0.5       			& 0.5 		& 0.1066\\												
			%\hline
		%\end{tabular}
	%\label{table:optimal0}
%\end{table}

\section{Conclusion}\label{sect:Conclusion}
We have studied a multihypothesis social learning problem in which an agent makes a decision with the help of a chosen expert. We have considered a general framework that allows the agent and experts to have different loss functions (biases), and different decision spaces. We have characterized the loss exponent of the agent in terms of the chosen expert's probability error exponent, which allows us to choose the asymptotically optimal expert as well as the agent's policy. We have shown that if the experts have the same decision space as the agent, then it is not necessarily optimal for the agent to choose an expert with the same loss function as itself. Moreover, if the agent is hypothesis-loss neutral, then it ignores any expert with a decision space smaller than the number of hypotheses. 

%<*tag:limitation1>
The results in this paper are limited by our setup and assumptions, which however allows us to obtain analytical characterizations of the asymptotically optimal policies for the agent and experts, and an asymptotically optimal expert choice. In the regime of a finite number of private observations, these policies are in general sub-optimal. Finding the exact optimal policies in this regime is however analytically difficult, and the asymptotically optimal policies can be utilized when the numbers of private observations are large.
%</tag:limitation1>

%<*tag:limitation2>
In this paper, we have assumed that the agent knows the experts' loss decay rates, which may not be valid in some practical scenarios. For example, expert opinions may very well depend on the mood of the expert at the time the opinion is publicized. Therefore, it is of interest to consider minimax loss exponents in which the expert's loss function or observation probability distributions are drawn from uncertainty classes. Minimax decentralized hypothesis testing has been studied in \cite{VeeBasPoo:94}, and in our recent work \cite{HoTayQuek:C14} in which we consider robust social learning in a tandem network. However, additional research in minimax decentralized hypothesis testing is required to address more complex network architectures and applications like that considered in this paper. 
%</tag:limitation2>

\appendices

\section{Mathematical Preliminaries}\label{appendix:Mathematical}

In this appendix, we briefly review some basic definitions and a result from large deviations theory. The reader is referred to \cite{DemZei:98,Str:93} for details. The notation $\trace{A}$ denotes the trace of the matrix $A$, $\nabla_t f(t)$ is the gradient of $f(t)$ w.r.t.\ the vector $t$, and $\nabla_t^2 f(t)$ is the Hessian w.r.t.\ $t$ of the function $f(t)$.

Let $\mathcal{X}$ be a Polish space. The function $\Phi : \mathcal{X} \mapsto [0,\infty]$ is said to be a good rate function if $\Phi$ is lower semicontinuous and has compact level sets $\{x \in\mathcal{X}: \Phi(x) \leq c\}$ for all $c \geq 0$. Let $(\P_n)_{n\geq1}$ be a sequence of probability measures on $\mathcal{X}$. This sequence of probability measures is said to satisfy a \emph{large deviation principle} (LDP) if for some good rate function $\Phi$, and for all closed sets $C \subset \mathcal{X}$, we have
\begin{align*}
\limsup_{n\to\infty} \ofrac{n}\log \P_n(C) \leq -\inf_{x\in C} \Phi(x),
\end{align*}
and for all open sets $O \subset\mathcal{X}$, we have
\begin{align*}
\liminf_{n\to\infty} \ofrac{n}\log \P_n(O) \geq -\inf_{x\in O} \Phi(x).
\end{align*}
A sequence of random variables $(Z_n)_{n\geq1}$ is said to satisfy a LDP if the sequence of marginal distributions $\P_n(\cdot) = \P(Z_n \in \cdot)$ satisfies a LDP. The celebrated G\"artner-Ellis Theorem \cite{DemZei:98} provides sufficient conditions for a sequence of random variables to satisfy a LDP. In the following, we present a partial version of the G\"artner-Ellis Theorem
\begin{Theorem_A}
Let $(Z_n)_{n\geq 1}$ be a sequence of $\Real^M$-valued random variables, so that
\begin{align*}
\varphi(t) = \lim_{n\to\infty}\ofrac{n}\log \EE{\exp(\ip{t}{Z_n})}
\end{align*}
exists. Let 
\begin{align*}
\Phi^*(z) = \sup_t \left\{ \ip{t}{z} - \varphi(t) \right\}
\end{align*}
be the Fenchel-Legendre transform of $\varphi$. Then, under the assumptions in Theorem 2.3.6 of \cite{DemZei:98}, the LDP holds with good rate function $\Phi^*$.
\end{Theorem_A}

When not all the conditions required by the G\"artner-Ellis Theorem hold (as is the case in some of our proofs), we instead make use of a \emph{uniform} lower bound given by the following lemma, which is a generalization of Theorem 1.3.13 of \cite{Str:93}. The proof is omitted here for brevity. 

\begin{Lemma_A}\label{lemma:LDP_lowerbound}
Let $Z_1,Z_2,\ldots,Z_n$ be independent $\Real^M$-valued random variables, with $x[\gamma]$ fraction of them having distribution $\P^\gamma$, for each $\gamma \in \Gamma$. Let $Z^\gamma$ have distribution $\P^\gamma$, 
\begin{align*}
\varphi(t) = \sum_{\gamma\in\Gamma} x[\gamma] \log \EE{\exp(\ip{t}{Z^\gamma})},
\end{align*}
and
\begin{align*}
\Phi^*(z) = \sup_t \left\{ \ip{t}{z} - \varphi(t) \right\}.
\end{align*}
Suppose that $\EE{\exp(\ip{t}{Z^\gamma})} < \infty$ for all $t\in\Real^M$, then for $t_z$ such that $\nabla_t\varphi(t_z) = z$, and any $\epsilon > 0$, we have
\begin{dmath}[label={A:lowerbound}]
\ofrac{n}\log \P\left( \ofrac{n} \sum_{l=1}^n Z_l \in B_\epsilon(z) \right) 
\geq - \Phi^*(z) - \norm{t_z}\epsilon + \ofrac{n}\log\left(1 \hiderel{-} \ofrac{n\epsilon^2} \trace{\nabla_t^2 \varphi(t_z)} \right).
\end{dmath}
\end{Lemma_A}

In this paper, we apply Lemma \ref{lemma:LDP_lowerbound} in cases where $\varphi$ corresponds to $\varphi_m$ in \eqref{logmomentgen}. We will often need to further lower bound \eqref{A:lowerbound} by finding an upper bound for $\nabla_t^2 \varphi_m(t_z)$ that does not depend on $n$ or the particular policy adopted by an agent. Assumption \ref{assumption:xi} implies an upper bound for $\nabla_t^2 \varphi_m$, as shown in the following elementary lemma.

\begin{Lemma_A}\label{lemma:bounded_trace}
Suppose that Assumption \ref{assumption:xi} holds. Then there exists a non-decreasing function $G(r)$, finite for each $r\geq0$, such that for all $t \in B_r(0)$, and all $\gamma\in\bigcup_k \Gamma_k$, we have for all $m\geq0$,
\begin{align}\label{trace_bound}
0\leq \trace{\nabla_t^2 \xi_m(\gamma,t)} \leq G(r).
\end{align}
\end{Lemma_A}
\begin{IEEEproof}
The lower inequality in \eqref{trace_bound} holds because $\xi_m(\gamma,t)$ is convex in $t$ for each $\gamma$ (see Lemma 2.2.31 of \cite{DemZei:98}). Furthermore, we can define 
\begin{align*}
G(r) = \max_{\substack{t\in B_r(0) \\ \gamma\in\Gamma}} \trace{\nabla_t^2 \xi_m(\gamma,t)},
\end{align*}
which is finite because of Assumption \ref{assumption:xi}, and the fact that $\trace{\nabla_t^2 \xi_m(\gamma,t)}$ is continuous on the compact set $B_r(0)$. The proof of the lemma is now complete.
\end{IEEEproof}

\section{Proofs of Main Results}\label{appendix:proofs}

\subsection{Proof of Lemma \ref{lemma:f_uniform}}
Let $\epsilon$ be a positive number. From Assumption \ref{assumpt:loss}\eqref{it:c_k}, for $n$ sufficiently large, we have $(\log M)/n \leq \epsilon /2$, $\min_m (\log \pi_m)/n \geq -\epsilon/2$,  and 
\begin{align*}
\left| \ofrac{n} \log C_k(m,d,n) + c_k(m,d) \right| \leq \frac{\epsilon}{2},
\end{align*}
for all $m \in [0,M-1]$. This implies that for all $z \in \Real^M$, we have
\begin{align*}
\tilde{g}_k(z,d,n) &\leq \max_{m} \{\ofrac{n} \log C_k(m,d,n) + z[m]\} + \ofrac{n} \log M \\
&\leq \max_{m} \{-c_k(m,d) + z[m]\} + \epsilon \\
& = \tilde{f}_k(z,d) + \epsilon,
\end{align*} 
and
\begin{align*}
\hspace{-2pt}\tilde{g}_k(z,d,n) &\geq \max_m \left\{\ofrac{n} \log C_k(m,d,n) + z[m] \right\} + \min_m \frac{\log \pi_m}{n} \\
&\geq   \max_{m} \{-c_k(m,d) + z[m]\} - \epsilon\\
& = \tilde{f}_k(z,d) - \epsilon,
\end{align*}
which shows that $\tilde{g}_k(z, d, n) \to \tilde{f}_k(z, d)$ uniformly in $z$. Since the minimum of a finite set of uniformly convergent functions is also uniformly convergent, the lemma follows.

\subsection{Proof of Theorem \ref{theorem:expert}}

We prove claim \eqref{it:expert_opt} of Theorem \ref{theorem:expert} by first deriving a lower bound for the loss exponent of agent $k$, and showing that this bound is achievable. We have for any $\epsilon > 0$, and $n_k$ sufficiently large,
\begin{align}
&\ofrac{n_k} \log \EE{C_k(H, D_k, n_k)} \nonumber\\
&= \ofrac{n_k} \log \sum_{m=0}^{M-1} \sum_{d=0}^{d_k-1} \pi_m C_k(m,d,n_k) \P_m(D_k = d) \nonumber\\
& \geq \max_{\substack{0\leq m \leq M-1 \\ 0\leq d \leq d_k-1}} \left\{ \ofrac{n_k} \log C_k(m,d,n_k) + \ofrac{n_k} \log\P_m(D_k = d) \right\}\nonumber\\
&\qquad\qquad +  \min_m\ofrac{n_k} \log \pi_m \nonumber\\
& \geq \max_{\substack{0\leq m \leq M-1 \\ 0\leq d \leq d_k-1}} \left\{ \ofrac{n_k} \log\P_m(D_k = d) -c_k(m,d)\right\} - \epsilon, \label{Ck_lowerbound}
\end{align}
which can be further lower bounded by lower bounds on the probability exponents. For each $z \in \Real^{M-1}$, let $t_z$ be the solution to the equation
\begin{align*}
\nabla_t \varphi_i(t,x_{k,n_k}) = z,
\end{align*}
if the solution exists. For each $r > 0$, let $H_r = \{ z \in \Real^{M-1} : t_z \textrm{ exists, } \norm{t_z} \leq r\}$, and $A_k(d,r,\epsilon) = \{z \in \Real^{M-1}: f_k(z^0, d) < -\epsilon/2,\textrm{ where } z^0 = (0, z)\} \cap H_r$. From \eqref{probD} and \eqref{barZ}, we then have for any $u < -\epsilon$,
\begin{align}
&\ofrac{n_k} \log\P_m(D_k = d)\nonumber\\
& = \ofrac{n_k} \log\P_m(g_k(\bar{Z}_{n_k}^0(x_{k,n_k}), d, n_k) < 0) \label{eqn:Df}\\
& \geq \ofrac{n_k} \log\P_m(g_k(\bar{Z}_{n_k}^0(x_{k,n_k}), d, n_k) \in [u-\epsilon,u+\epsilon]) \nonumber\\
& \geq \ofrac{n_k} \log\P_m(f_k(\bar{Z}_{n_k}^0(x_{k,n_k}), d) \in [u-\frac{\epsilon}{2},u+\frac{\epsilon}{2}]) \nonumber
\end{align}
where the last inequality follows from Lemma \ref{lemma:f_uniform} for $n_k$ sufficiently large.\footnote{Note that Assumption \ref{assumpt:unambiguous} is required in \eqref{eqn:Df}, without which we need to replace $\P_m(g_k(\bar{Z}_{n_k}^0(x_{k,n_k}), d, n_k) < 0)$ with $\max\{\P_m(g_k(\bar{Z}_{n_k}^0(x_{k,n_k}), d, n_k) < 0), \P_m(g_k(\bar{Z}_{n_k}^0(x_{k,n_k}), d, n_k) = 0)\}$ if expert $k$ uses randomization to produce its final decision. The second term in the maximization unfortunately cannot be characterized using our existing approach.} Since $u < -\epsilon$ is arbitrary, we obtain for $n_k$ sufficiently large,
\begin{align}
&\ofrac{n_k} \log\P_m(D_k = d) \nonumber\\
& \geq \sup_{z : B_\epsilon(z) \subset A_k(d,r,\epsilon)} \ofrac{n_k} \log\P_m(\bar{Z}_{n_k}(x_{k,n_k}) \in B_{\epsilon}(z)),\label{lowerbound1}
\end{align}
where $B_\epsilon(z)$ is an open sphere of radius $\epsilon$ around $z$. From Lemma \ref{lemma:LDP_lowerbound}, we can further lower bound the right hand side of \eqref{lowerbound1} to obtain
\begin{align}
&\ofrac{n_k} \log\P_m(D_k = d)  \nonumber\\
&\geq - \inf_{z : B_\epsilon(z) \subset A_k(d,r,\epsilon)} \bigg\{ \Phi_m^*(z, x_{k,n_k}) + \norm{t_z}\epsilon \nonumber\\
&\qquad - \ofrac{n_k}\log\left(1 - \frac{1}{n_k\epsilon^2} \trace{\nabla_t^2 \varphi_i(t_z,x_{k,n_k})}\right) \bigg\}\nonumber\\
&\geq - \inf_{z : B_\epsilon(z) \subset A_k(d,r,\epsilon)} \Phi_m^*(z, x_{k,n_k}) + r\epsilon \nonumber\\
&\qquad - \ofrac{n_k}\log\left(1 - \frac{1}{n_k\epsilon^2} G(r)\right),\label{lowerbound2}
\end{align}
where the last inequality follows from Lemma \ref{lemma:bounded_trace}. Combining \eqref{lowerbound2} with \eqref{Ck_lowerbound}, and letting $n_k\to\infty$, and then taking $r = 1/\sqrt{\epsilon}$ and $\epsilon \to 0$, we have
\begin{align}
\liminf_{n_k \to\infty} \ofrac{n_k} \log \EE{C_k(H, D_k, n_k)}
& \geq - \sup_{x \in \mathbb{S}(\Gamma_k)} I_k(x). \label{Ck_lowerbound2}
\end{align}

Since $\Phi_m^*(z,x)$ is continuous in $x$, $\inf_{z \in A_k(d)} \Phi_m^*(z, x)$ is upper semi-continuous in $x$, and $I_k(x)$ is upper semi-continuous in $x$. In addition, $\mathbb{S}(\Gamma_k)$ is compact, therefore the supremum over $x$ on the right hand side of \eqref{Ck_lowerbound2} is a maximization. Consider the policy $x_k^* = \arg\max_{x \in \mathbb{S}(\Gamma_k)} I_k(x)$. For each $n_k \geq 1$, let agent $k$ use the policy $x_{k,n_k}$ where $x_{k,n_k}[\gamma] = \floor{x_k^*[\gamma] n_k}/n_k$ for all $\gamma\in\Gamma_k$, and if $x_{k,n_k}[\gamma]$ do not sum to 1 over $\gamma\in\Gamma_k$, we simply choose the remaining private observations from an arbitrary distribution, and ignore them when making the decision for agent $k$. We have $x_{k,n_k} \to x_k^*$ as $n_k\to\infty$. We first show a simple lemma.
\begin{Lemma_A}\label{lemma:LDP_Z}
Suppose that Assumption \ref{assumption:xi} holds, agent $k$ adopts the policy $x_{k,n_k}=(x_{k,n_k}[\gamma])_{\gamma\in\Gamma_k}$ when it has access to $n_k$ private observations, and $x_{k,n_k} \to x_k$ as $n_k \to \infty$. Then, the sequence of random variables $(\bar{Z}_{n_k}(x_{k,n_k}))_{n_k \geq 1}$ defined in \eqref{barZ} satisfies a LDP under hypothesis $H=m$, for every $m\in [0,M-1]$, with good rate function $\Phi_m^*(\cdot,x_k)$.
\end{Lemma_A}
\begin{IEEEproof}
We apply the G\"artner-Ellis Theorem \cite{DemZei:98} to prove the lemma. Let $Z_i = (\log \ell_{m0}(Y_k[i]))_{m=1}^{M-1}$. We have $n_k \bar{Z}_{n_k}(x_{k,n_k}) = \sum_{i=1}^{n_k} Z_i$ since $Z_i, i=1\ldots,n_k$ are independent. For every $t\in\Real^{M-1}$, we obtain
\begin{align*}
&\ofrac{n_k} \log \Ec{m}{\exp\left(\ip{n_k t}{\bar{Z}_{n_k}(x_{k,n_k})}\right)} \\
&=\ofrac{n_k} \log \Ec{m}{\exp\left(\Big\langle t, \sum_{i=1}^{n_k} Z_i \Big\rangle \right)} \\
&= \ofrac{n_k} \sum_{i=1}^{n_k} \log \Ec{m}{\exp\left(\ip{t}{Z_i}\right)} \\
&= \sum_{\gamma\in\Gamma_k} x_{k,n_k}[\gamma] \log \Ec{m}{\exp\left(\ip{t}{(\log\ell_{m0}^\gamma)_{m=1}^{M-1}}\right)} \\
& \to \varphi_m(t,x_{k}),
\end{align*}
as $n_k \to \infty$. From Assumption \ref{assumption:xi} and Lemma 2.3.9 of \cite{DemZei:98}, we have $\Phi_m^*(z,x_k)$ is a good rate function, and the lemma follows from the G\"artner-Ellis Theorem.
\end{IEEEproof}
From Lemma \ref{lemma:f_uniform}, we have for each $d \in [0,d_k-1]$,
\begin{align*}
\limsup_{n_k\to\infty} \sup_z |g_k(z,d,n_k) - f_k(z,d)| =0,
\end{align*}
and applying Theorem 4.2.23 of \cite{DemZei:98}, we obtain from \eqref{eqn:Df} and Lemma \ref{lemma:LDP_Z} that
\begin{align}\label{Dk_upperbound}
\limsup_{n_k\to\infty} \ofrac{n_k} \log \P_m(D_k = d) \leq - \inf_{z \in A_k(d)} \Phi_m^*(z,x_k^*).
\end{align}
We then have
\begin{align}
&\limsup_{n_k\to\infty} \ofrac{n_k} \log \EE{C_k(H, D_k, m_l)} \nonumber\\
&\leq \limsup_{n_k\to\infty} \max_{\substack{0\leq m \leq M-1 \\ 0\leq d \leq d_k-1}} \left\{\ofrac{n_k} \log\P_m(D_k = d) -c_k(m,d) \right\} \nonumber\\
&\leq  -I_k(x_k^*), \label{Ck_upperbound}
\end{align}
where the last inequality follows from \eqref{Dk_upperbound}. Finally, \eqref{Ck_lowerbound2} together with \eqref{Ck_upperbound} gives us claim \eqref{it:expert_opt}. 

To show claim \eqref{it:policy_opt}, fix any $x_k^* \in \arg\max_{x \in \mathbb{S}(\Gamma_k)} I_k(x)$. We note that if there exists a subsequence of policies $(x_{k,m_l})_{l\geq 1}$ with $\lim_{l\to\infty}x_{k,m_l} = x_k$ and $I_k(x_k) < I_k(x_k^*)$, then using the same arguments that lead to \eqref{Ck_lowerbound2}, we have 
\begin{align*}
\liminf_{n_k \to\infty} \ofrac{n_k} \log \EE{C_k(H, D_k, n_k)}
& \geq -  I_k(x_k) > - I_k(x_k^*),
\end{align*}
a contradiction to \eqref{Ck_upperbound}. Therefore, each policy subsequence converges to some $x_k$ with $I_k(x_k) = I_k(x_k^*)$, and there is no loss in optimality if we restrict the sequence of policies to converge to $x_k^*$. 

Finally, to show claim \eqref{it:prob_exp}, we have from \eqref{lowerbound2} that
\begin{align}\label{Dk_lowerbound}
\liminf_{l\to\infty}\ofrac{n_k} \log\P_m(D_k = d) &\geq - \inf_{z \in A_k(d)} \Phi_m^*(z,x_k^*), 
\end{align}
since $\Phi_m^*(z,x)$ is continuous in $x$. Together with \eqref{Dk_upperbound}, the claim now follows, and the theorem is proved.

\subsection{Proof of Lemma \ref{lemma:Phi_properties}}\label{appendix:lemma:Phi_properties}
The non-negativity and convexity of $\Phi_m(z,x)$ follows from Lemma 2.2.31 of \cite{DemZei:98}. From Jensen's inequality, for any $t\in\Real^{M-1}$, we have
\begin{align*}
\varphi_m(t,x) 
&\geq \sum_{\gamma} x[\gamma] \Ec{m}{\ip{t}{Z^\gamma}} = \ip{t}{\tilde{z}_m(x)},
\end{align*}
which implies that $\Phi_m(\tilde{z},x) = 0$, and the lemma is proved.

\subsection{Proof of Theorem \ref{theorem:agent0}}

We first present a generalization of Theorem 5 of \cite{ShaGalBer:67a} (see also \cite{Tay:J12} for a slightly more updated version). The proof steps are similar to that in \cite{ShaGalBer:67a} and \cite{Tay:J12}, and are provided below for completeness.

\begin{Proposition_A}\label{prop:D0_lowerbound}
Suppose that Assumptions \ref{assumpt:loss} and  \ref{assumption:xi} hold, and agent $0$ adopts the opinion of agent $k \geq 1$ and policy $x_0$. Let
\begin{align*}
s_{ij}^*\hspace{-4pt} = \arg\hspace{-4pt} \max_{s\in [0,1]} \left\{\frac{s}{n_0} \log \frac{C_0(i,j,n_0)\P_i(D_k=d)}{C_0(j,i,n_0)\P_j(D_k=d)} - \Lambda_{ij}(s,x_0)\right\}
\end{align*}
where $\Lambda_{ij}(\cdot,\cdot)$ is as defined in \eqref{def:Lambdaij}. Then, for any $\epsilon > 0$, and any $d \in [0,d_k-1]$, there exists $n$ such that for all $n_0 \geq n$, we have for all $i\ne j$, 
\begin{align}
& \ofrac{n_0}\log \left\{ 
\begin{array}{cc}
\min_{j':j'\ne i} C_0(i,j',n_0)\P_i(D_0(k) \ne i, D_k = d) \\
+ C_0(j,i,n_0)\P_j(D_0(k)=i, D_k=d) 
\end{array}
\right\}\nonumber\\
& \geq (1-s_{ij}^*)\left(\ofrac{n_0}\log\P_i(D_k=d) - c_0(i)\right) \nonumber\\
& \qquad + s_{ij}^*\left(\ofrac{n_0}\log\P_j(D_k=d) - c_0(j) \right) 
+ \Lambda_{ij}(s_{ij}^*, x_0) - \epsilon.\label{D0_lowerbound}
\end{align}
\end{Proposition_A}
\begin{IEEEproof}
From Theorem 5 of \cite{ShaGalBer:67a} (or Proposition A.2 of \cite{Tay:J12}), we have for $i,j \in [0,M-1]$, with $j \neq i$, and every $s \in [0,1]$, either
\begin{dmath}[compact,label={D0_lowerbound1}]
\P_i(D_0(k) \hiderel{\ne} i \hiderel{\mid} D_k \hiderel{=} d) 
\geq \ofrac{4}\exp\left(n_0\Lambda_{ij}(s,x_0) \hiderel{-} s n_0\ddfrac{}{s}\Lambda_{ij}(s,x_0) 
- s\sqrt{2n_0 \ddfrac{^2}{s^2}\Lambda_{ij}(s,x_0)}\ \right),
\end{dmath}
or
\begin{dmath}[compact,label={D0_lowerbound2}]
\P_j(D_0(k) \hiderel{=} i \hiderel{\mid} D_k \hiderel{=} d) 
\geq \ofrac{4}\exp\left(n_0\Lambda_{ij}(s,x_0) \hiderel{+} (1 \hiderel{-} s)n_0\ddfrac{}{s}\Lambda_{ij}(s,x_0) 
- (1-s)\sqrt{2n_0\ddfrac{^2}{s^2}\Lambda_{ij}(s,x_0)}\ \right).
\end{dmath}
If $s_{ij}^* \in (0,1)$, we have
\begin{align*}
\ddfrac{}{s}\Lambda_{ij}(s_{ij}^*,x_0) = \frac{1}{n_0} \log \frac{C_0(i,j,n_0)\P_i(D_k=d)}{C_0(j,i,n_0)\P_j(D_k=d)},
\end{align*}
and using Lemma \ref{lemma:bounded_trace}, \eqref{D0_lowerbound1} and \eqref{D0_lowerbound2}, we obtain
\begin{align*}
& \ofrac{n_0}\log \left\{ 
\begin{array}{cc}
\min_{j': j'\ne i} C_0(i,j',n_0)\P_i(D_0(k) \ne i, D_k = d) \\
+ C_0(j,i,n_0)\P_j(D_0(k)=i, D_k=d) 
\end{array}
\right\}\\
& \geq \frac{1-s_{ij}^*}{n_0}\log (\min_{j': j'\ne i} C_0(i,j',n_0)\P_i(D_k=d))\\
& \quad\quad + \frac{s_{ij}^*}{n_0}\log (C_0(j,i,n_0)\P_j(D_k=d)) \\ 
& \quad\quad + \Lambda_{ij}(s_{ij}^*, x_0) - \ofrac{n_0}\log 2 - \sqrt{\frac{2G(2)}{n_0}}\\
& \geq (1-s_{ij}^*)\left(\ofrac{n_0}\log\P_i(D_k=d) - c_0(i)\right) \\
&\qquad + s_{ij}^*\left(\ofrac{n_0}\log\P_j(D_k=d) - c_0(j)\right) + \Lambda_{ij}(s_{ij}^*, x_0) - \epsilon,
\end{align*}
where the last inequality follows from Assumption \ref{assumpt:loss} for $n_0$ sufficiently large. On the other hand, if $s_{ij}^* =0$, we have 
\begin{align*}
\ddfrac{}{s}\Lambda_{ij}(0,x_0)\geq \frac{1}{n_0} \log \frac{C_0(i,j,n_0)\P_i(D_k=d)}{C_0(j,i,n_0)\P_j(D_k=d)},
\end{align*}
since $\Lambda_{ij}(s,x_0)$ is convex $s$. The inequality \eqref{D0_lowerbound} then holds trivially. A similar argument holds for $s_{ij}^*=1$, and the proposition is proved.
\end{IEEEproof}

We next proceed to prove Theorem \ref{theorem:agent0}. Suppose that agent $0$ adopts the policy $x_0$. For any $\epsilon > 0$, and for $n_0$ sufficiently large, we have
\begin{align}
&\ofrac{n_0}\log \EE{C_0(H,D_0(k),n_0)} \nonumber\\
& = \ofrac{n_0}\log \sum_{d=0}^{d_k-1} \sum_{i\ne j} \pi_i C_0(i,j,n_0)\P_i(D_0(k)=j, D_k=d) \nonumber\\
& \geq\hspace{-8pt} \max_{\substack{i\ne j \\ 0\leq d \leq d_k-1}} \hspace{-8pt}\ofrac{n_0}\log\left\{ 
\begin{array}{cc}
\min_{j': j'\ne i} C_0(i,j',n_0) \\
\qquad \cdot \P_i(D_0(k) \ne i, D_k = d)\\
+ C_0(j,i,n_0)\P_j(D_0(k)=i, D_k=d)
\end{array}
\right\} \nonumber\\ 
&\qquad - \ofrac{n_0}\log 2 + \min_m \ofrac{n_0} \log \pi_m \nonumber\\
& \geq \max_{\substack{i\ne j \\ 0\leq d \leq d_k-1}} \min_{s\in [0,1]} \Big\{ (1-s)\left(\ofrac{n_0}\log\P_i(D_k=d) - c_0(i)\right) \nonumber\\ 
& + s\left(\ofrac{n_0}\log\P_j(D_k=d)- c_0(j)\right)
 + \Lambda_{ij}(s, x_0) \Big\} - \epsilon,\label{C0_lowerbound1}%
\end{align}%
where the last inequality follows from Proposition \ref{prop:D0_lowerbound}. By letting $n_0\to\infty$ and $\epsilon\to0$ in \eqref{C0_lowerbound1}, we obtain 
\begin{align}
&\limsup_{n_0\to\infty} \ofrac{n_0}\log \EE{C_0(H,D_0(k),n_0)} \nonumber\\
& \geq \hspace{-10pt}\max_{\substack{i\ne j \\ 0\leq d \leq d_k-1}}\hspace{-10pt} \min_{s\in [0,1]} \Big\{ (1-s)\left(\lim_{n_0\to\infty}\ofrac{n_0}\log\P_i(D_k=d) - c_0(i)\right) \nonumber\\ 
& \quad + s\left(\lim_{n_0\to\infty}\ofrac{n_0}\log\P_j(D_k=d)- c_0(j)\right)
 + \Lambda_{ij}(s, x_0) \Big\}, \label{C0_lowerbound1_limit}
\end{align}
and the lower bound \eqref{C0_lowerbound} follows from Theorem \ref{theorem:expert}\eqref{it:prob_exp}.

We next show that there exists a decision rule for agent $0$ that achieves $\cE_0(x_0)$ in \eqref{C0_lowerbound}. Given $D_k=d$, consider the following rule to differentiate between hypotheses $H=i$ and $H=j$ for $i\ne j$: declare $H=i$ iff $\ofrac{n_0}\log \ell_{ji}(Y_0[1:n_0]) \leq h_{ji}$, 
where $h_{ji} =  -q_k\inf_{z \in A_k(d)}\Phi_i^*(z, x_k^*) + q_k\inf_{z \in A_k(d)}\Phi_j^*(z, x_k^*) - c_0(i) + c_0(j)$.

By a simple generalization of Cram\'er's Theorem\footnote{Cram\'er's Theorem applies to independent and identically distributed (i.i.d.) observations. The private observations $Y_0[1:n_0]$ are not i.i.d., but are independent and can be divided into groups of i.i.d.\ observations.} \cite{DemZei:98}, and Theorem \ref{theorem:expert}\eqref{it:prob_exp}, we have for every $\epsilon >0$ and all $n_0$ sufficiently large,
%\begin{align*}
%&\ofrac{n_0}\log (C_0(i,j,n_0)\P_i(D_0(k)=j, D_k=d)) \\
%&= \ofrac{n_0}\log C_0(i,j,n_0)+ \ofrac{n_0}\log \P_i(D_k=d) \\
%&\qquad + \ofrac{n_0}\log \P_i(D_0(k)=j\mid D_k=d) \\
%&\leq -c_0(i) - q_k\inf_{z \in A_k(d)}\Phi_i^*(z, x_k^*) \\
%&\qquad - \max_{s\in[0,1]} \left\{ sh_{ji} - \Lambda_{ij}(s,x_0)\right\} + \epsilon \\
%& \leq -\max_{s\in [0,1]} \bigg\{  (1-s)\left(q_k\inf_{z \in A_k(d)} \Phi_i^*(z, x_k^*) + c_0(i)\right) \nonumber\\
%&\quad\quad + s\left(q_k\inf_{z \in A_k(d)} \Phi_j^*(z, x_k^*)+c_0(j)\right) - \Lambda_{ij}(s,x_0) \bigg\} + \epsilon,
%\end{align*}
\begin{dmath*}
\ofrac{n_0}\log (C_0(i,j,n_0)\P_i(D_0(k)\hiderel{=}j, D_k\hiderel{=}d)) 
= \ofrac{n_0}\log C_0(i,j,n_0)+ \ofrac{n_0}\log \P_i(D_k\hiderel{=}d) 
+ \ofrac{n_0}\log \P_i(D_0(k)\hiderel{=}j\hiderel{\mid} D_k\hiderel{=}d) 
\leq -c_0(i) - q_k\inf_{z \in A_k(d)}\Phi_i^*(z, x_k^*) 
- \max_{s\in[0,1]} \left\{ sh_{ji} - \Lambda_{ij}(s,x_0)\right\} + \epsilon 
\leq -\max_{s\in [0,1]} \bigg\{  (1-s)\left(q_k\inf_{z \in A_k(d)} \Phi_i^*(z, x_k^*) + c_0(i)\right) 
+ s\left(q_k\inf_{z \in A_k(d)} \Phi_j^*(z, x_k^*)+c_0(j)\right) - \Lambda_{ij}(s,x_0) \bigg\} + \epsilon,
\end{dmath*}
from which we obtain
%\begin{align*}
%&\ofrac{n_0}\log \EE{C_0(H,D_0(k),n_0)} \\
%& \leq \max_{\substack{i\ne j \\ 0\leq d \leq d_k-1}} \ofrac{n_0}\log\left(C_0(i,j,n_0)\P_i(D_0(k) =j, D_k = d)\right) \\
%&\qquad + \ofrac{n_0} \log M \\
%& \leq -\cE_0(k,x_0) + \epsilon.
%\end{align*}
\begin{dmath*}
\ofrac{n_0}\log \EE{C_0(H,D_0(k),n_0)} 
\leq \max_{\substack{i\ne j \\ 0\leq d \leq d_k-1}} \ofrac{n_0}\log\left(C_0(i,j,n_0)\P_i(D_0(k) \hiderel{=}j, D_k \hiderel{=} d)\right) + \ofrac{n_0} \log M
\leq -\cE_0(k,x_0) + \epsilon.
\end{dmath*}
By taking $n_0\to\infty$ and $\epsilon\to 0$, we obtain the theorem by maximizing $\cE_0(x_0)$ over all policies $x_0$. The proof is now complete.

\subsection{Proof of Theorem \ref{theorem:expert_choice}}
The first part of the theorem is a direct consequence of Theorem \ref{theorem:agent0}. From Lemma \ref{lemma:Phi_properties}, we have $\inf_{z \in A_k(d)} \Phi_i^*(z, x_k^*) \geq 0$ for any $i\in[0,M-1]$, $k\geq 1$, $d\in [0,d_k-1]$, and policy $x_k^*$. Furthermore, Assumption \ref{assumpt:loss} implies that $c_0(i) \geq 0$ for all $i\in[0,M-1]$. Therefore, the inequalities \eqref{Eineq1} and \eqref{Eineq2} follow from \eqref{C0_lowerbound}, \eqref{E0}, \eqref{EB}, and \eqref{EkB}, and the proof is complete.

\subsection{Proof of Proposition \ref{prop:expert_necessary}}
Suppose that agent $k$ adopts the policy $x_k$. From the pigeonhole principle, if $d_k < M$, there exists a region $A_k(d)$ in which both $\Phi_i(\cdot,x_k)$ and $\Phi_j(\cdot,x_k)$ achieve their minimum value of 0, for some $i\ne j$. Since for any $(i',j')$, we have
\begin{dmath*}
\max_{s\in [0,1]} \left\{  (1-s)\left(q_k\inf_{z \in A_k(d)} \Phi_{i'}^*(z, x_k^*) + c_0(i')\right)
+ s\left(q_k\inf_{z \in A_k(d)} \Phi_{j'}^*(z, x_k^*) \hiderel{+} c_0(j')\right) \hiderel{-} \Lambda_{i'j'}(s,x_0) \right\}
\geq \max_{s\in [0,1]} (-\oLambda_{i'j'}(s,x_0,c_0))
= -\min_{s\in [0,1]} \oLambda_{ij}(s,x_0,c_0),
\end{dmath*}
we obtain from \eqref{C0_lowerbound} and \eqref{E0}, 
\begin{align*}
&\cE_0(k,x_0) = -\min_{s\in [0,1]} \oLambda_{ij}(s,x_0,c_0) = \cE_{0,B}(k,x_0),
\end{align*}
and the proposition is proved.

\subsection{Proof of Proposition \ref{prop:bestexpertloss}}
Since the proof is similar to that of Theorem \ref{theorem:agent0}, we provide only an outline here. From the proposition assumptions, we have for every $j \in [0,M-1]$, $\P_j(D_k=j)$ is bounded away from zero, i.e., $\lim_{n_k\to\infty}(1/n_k)\log \P_j(D_k=j) = 0$ because otherwise the expected loss of agent $k$ can be decreased. Therefore, we have $\inf_{z\in A_k(j)} \Phi_j^*(z,x_k) = 0$. By comparing \eqref{E0} (with $x_0$ replaced by $x_k$) and \eqref{Ik}, and using an inductive argument, we have $\inf_{z\in A_k(i)} \Phi_j^*(z,x_k) = \Lambda_{ji}^*(c_k(i)-c_k(j), x_k)$, and \eqref{sp_bestlossexp} follows. 

We next show the second part of the proposition. Suppose that for some $i\ne j$, we have $c_0(i)=c_0(j)$ and \eqref{01condition} holds. Then from \eqref{eqn:oLambda}, we have $\oLambda_{ij}(s,x_0,c_0)=\oLambda_{ji}(s,x_0,c_0)$ for all $s\in[0,1]$. Let $g_{ij}(\Delta) = \max_{s\in [0,1]} \{ sq_k \Lambda^*_{ji}(\Delta,x_k) - \oLambda_{ij}(s,x_0,c_0) \}$. Note that $g_{ij}(\Delta)$ is non-decreasing in $\Delta$. Therefore, since $\oLambda_{ij}(s,x_0,c_0)$ is symmetrical on $s\in[0,1]$, we have $\min(g_{ij}(\Delta), g_{ji}(-\Delta))$ is maximized if $\Lambda^*_{ji}(\Delta,x_k)=\Lambda^*_{ij}(-\Delta,x_k)$, which holds if $\Delta=0$. Take $\Delta = c_k(i)-c_k(j)$, and the proposition follows.

\section{Characterization of Asymptotic Decision Regions}\label{appendix:Ak_properties}

In this appendix, we give a characterization for the asymptotic decision region $A_k(d)$ for an agent $k$, and $d \in [0,d_k-1]$. For $i, j \in [0,M-1]$ and $p, q\in [0,d_k]$, define the halfspace
\begin{align}\label{half-space}
B_k(i,p,j,q) = \big\{ & z=(z[m])_{1\leq m\leq M-1}\in \Real^{M-1}: \nonumber\\
& z[i] - z[j] \geq c_k(i,p) - c_k(j,q)\big\},
\end{align}
where $z[0] = 0$. For each $p \in [0,d_k-1]$, let $m_p \in [0,M-1]$ be a chosen corresponding index. From \eqref{tfk}, we have $z \in \cap_{i \ne m} B_k(m_p,p,i,p)$ iff $\tilde{f}_k(z,p) = z[m_p]-c_k(m_p,p)$. 
%Therefore, 
%\begin{align*}
%z \in \bigcap_{p=0}^{d_k-1} \bigcap_{i \ne m_p} B_k(m_p,p,i,p)
%\end{align*}
%iff $\tilde{f}_k(z,p) = z[m_p]-c_k(m_p,p)$ for all $p \in [0,d_k-1]$. 

Consider a $z\in\Real^{M-1}$ such that $f(z,d) < 0$. Then, there exists a sequence $(m_p)_{0\leq p \leq d_k-1} \in [0,M-1]^{d_k}$ such that $\tilde{f}_k(z,p) = z[m_p]-c_k(m_p,p)$ for all $p\in[0,d_k-1]$ and $\tilde{f}_k(z,p) - \tilde{f}_k(z,d) = z[m_p] - z[m_d] - c_k(m_p,p) + c_k(m_d,d) > 0$ for all $p \ne d$, i.e., 
\begin{align*}
z \in & H_d((m_p)_{p=0}^{d_k-1}) \\
& \triangleq \bigcap_{p=0}^{d_k-1}\bigcap_{i \ne m_p} B_k(m_p,p,i,p)\bigcap_{p \ne d} B_k(m_p,p,m_{d}, d),
\end{align*}
where $H_d((m_p)_{p=0}^{d_k-1})$ is a polyhedron since it consists of intersections of halfspaces. On the other hand, if such a sequence $(m_p)_{p=0}^{d_k-1}$ exists, then $z \in A_k(d)$. Therefore, the set $A_k(d)$ is the union over all sequences $(m_p)_{p=0}^{d_k-1} \in [0,M-1]^{d_k}$ of the polyhedra $H_d((m_p)_{p=0}^{d_k-1})$.

%\bibliographystyle{IEEEtran}
%\bibliographystyle{plain}
%\bibliography{IEEEabrv,StringDefinitions,DecentDet,BibBooks,Tay,Social,Infection,IoT}


\begin{IEEEbiography}
[{\includegraphics[width=1in,height =1.25in,clip,keepaspectratio]{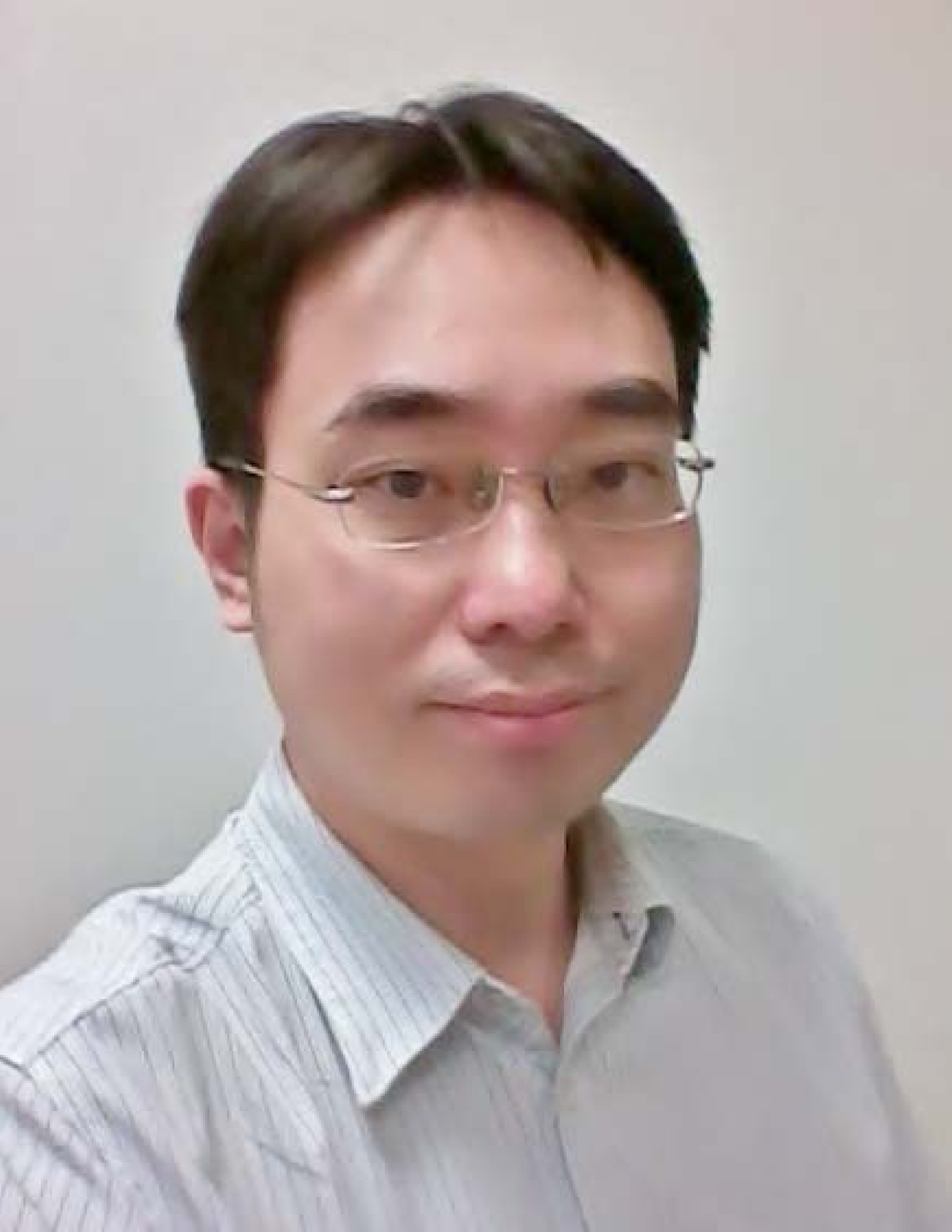}}]
{Wee Peng Tay} (S'06 M'08) received the B.S. degree in Electrical Engineering and Mathematics, and the M.S. degree in Electrical Engineering from Stanford University, Stanford, CA, USA, in 2002. He received the Ph.D. degree in Electrical Engineering and Computer Science from the Massachusetts Institute of Technology, Cambridge, MA, USA, in 2008. He is currently an Assistant Professor in the School of Electrical and Electronic Engineering at Nanyang Technological University, Singapore. His research interests include distributed detection and estimation, distributed signal processing, sensor networks, social networks, information theory, and applied probability.

Dr. Tay received the Singapore Technologies Scholarship in 1998, the Stanford University President's Award in 1999, and the Frederick Emmons Terman Engineering Scholastic Award in 2002. He is the coauthor of the best student paper award at the 46th Asilomar conference on Signals, Systems, and Computers. He is currently serving as the chair of DSNIG in IEEE MMTC, and has served as a technical program committee member for various international conferences.
\end{IEEEbiography}

\end{document}